\documentclass[aps,twocolumn]{revtex4}
\usepackage{hyperref}
\usepackage{graphicx}
\usepackage{subfigure}
\usepackage{dcolumn}
\usepackage{ae} 
\usepackage{float}  
\usepackage{amssymb}
\usepackage{mathtools}
\usepackage{amsmath}
\usepackage{epsfig}
\usepackage{epstopdf}
\usepackage[dvipsnames]{xcolor}
\usepackage{natbib}

\begin{document}

\title{Magneto-optical properties of Au upon the injection of hot spin-polarized electrons across Fe/Au(001) interfaces}

\author{A.~Alekhin$^{1}$, I.~Razdolski$^{1,2}$, M.~Berritta$^3$, D. B{\"u}rstel$^4$, V. Temnov$^{1}$, D.~Diesing$^4$, U.~Bovensiepen$^{5}$, G. Woltersdorf$^6$,  P.~M.~Oppeneer$^3$, and A.~Melnikov$^{2,6}$}

\affiliation{$^1$Institute of Molecules and Materials of Le Mans, CNRS UMR 6283, 72085 Le Mans, France}

\affiliation{$^2$Fritz Haber Institute of the Max Planck Society, 14195 Berlin, Germany}

\affiliation{$^3$Department of Physics and Astronomy, Uppsala University, Box 516, SE-75120 Uppsala, Sweden}

\affiliation{$^4$Faculty of Chemistry, University of Duisburg-Essen, Universit{\"a}tstra{\ss}e 5, Essen 45141, Germany}

\affiliation{$^5$Faculty of Physics and Center for Nanointegration (CENIDE), University of Duisburg-Essen, Lotharstra{\ss}e 1, Duisburg 47057, Germany}

\affiliation{$^6$Institute of Physics, Martin Luther University Halle-Wittenberg, Von-Danckelmann-Platz 3, 06120 Halle, Germany}

\begin{abstract}
We demonstrate a novel method for the excitation of sizable magneto-optical effects in Au by means of the laser-induced injection of hot spin-polarized electrons in Au/Fe/MgO(001) heterostructures. It is based on the energy- and spin-dependent electron transmittance of Fe/Au interface which acts as a spin filter for non-thermalized electrons optically excited in Fe. We show that after crossing the interface, majority electrons propagate through the Au layer with the velocity on the order of 1~nm/fs (close to the Fermi velocity) and the decay length on the order of 100 nm. Featuring ultrafast functionality and requiring no strong external magnetic fields, spin injection results in a distinct magneto-optical response of Au. We develop a formalism based on the phase of the transient complex MOKE response and demonstrate its robustness in a plethora of experimental and theoretical MOKE studies on Au, including our {\it ab initio} calculations. Our work introduces a flexible tool to manipulate magneto-optical properties of metals on the femtosecond timescale that holds high potential for active magneto-photonics, plasmonics, and spintronics.
\end{abstract}

\keywords{spin current, time-resolved MOKE, nonlinear magneto-optical spectroscopy, Drude model}

\maketitle

\section{Introduction}

Development of femtosecond laser technology has enabled vast possibilities for driving and monitoring optical properties of media on ultrashort timescales. Enjoying high potential for the magneto-optical recording applications~\cite{Kirilyuk2010, Kirilyuk2013}, ultrafast opto-magnetism routinely employs  magneto-optical response as a tool for the observation of magnetic states in a photo-excited medium. Initiated by the discovery of ultrafast laser-induced demagnetization~\cite{Beaurepaire1996}, ultrafast spin dynamics now considers both local and non-local electronic processes~\cite{Hohlfeld2000, Walowski2016, Razdolski2017b, Wieczorek2015}. A concept of superdiffusive spin transport across metallic layers and interfaces between them as a relevant mechanism for non-local spin dynamics in heterostructures was suggested~\cite{Battiato2010}, showing a good agreement with the experimental observations~\cite{Malinowski2008, Melnikov2011, Eschenlohr2013, Turgut2013}.

In particular, the injection of optically excited spin-polarized hot carriers from a ferromagnetic Fe film into an adjacent Au layer was evidenced using a magnetization-sensitive nonlinear-optical technique~\cite{Melnikov2011, Alekhin2017}. The lack of \emph{d}-states in the majority sub-band of Fe above the Fermi energy results in a good matching of the \emph{s-p} wave functions in both Fe and Au. This leads to a high transmittance of the Fe/Au interface for the majority electrons whereas for the minority ones (as well as for the majority electrons at lower energies) the average transmittance is low due to the large density of d-states in Fe.
Since the laser pulse promotes majority electrons in Fe into the states well above the Fermi energy while leaving Au essentially unperturbed, a sizable flux of hot majority electrons across the interface in strongly excited non-thermal systems can be expected within the thermalization time of the electron distribution~\cite{Alekhin2017}. Noteworthy, this points to the generality of the spin filter  properties of the noble metal-ferromagnet interfaces for hot electrons. As such, unique properties of hot electrons in various metallic systems~\cite{Baida2011,Mukherjee2013, Reiner2017, Hartland2017} have been complemented with the ability to employ the noble metal-ferromagnet interfaces as an efficient spin filter, thereby opening the door to magneto-optical characterization of spin properties of buried interfaces.

In fact, featuring ultimate time resolution, a new type of optical probe~\cite{Alekhin2017} tackles the problem of experimental investigations of energy- and spin-dependent electronic transmittance of interfaces. Importantly, relying on the (long-range) hot electron spin transport, this \emph{non-local} technique is advantageous for probing the electronic and magnetic state of buried interfaces, as compared to the established \emph{local} optical methods. In turn, making use of the spin filter properties of the interface for the non-thermalized electrons excited in a ferromagnet, laser-induced spin injection holds high promise for ultrafast all-optical manipulation of the transient magnetic state and magneto-optical response of paramagnetic media.

Yet, the employment of time-resolved magneto-optical methods as a probe for magnetization requires an unconditional relation between the latter and the magnitude of the magneto-optical effects on all timescales. In a number of systems, this assumption has been shown to break down, often illustrated by the incommensurability of the transients of the real and imaginary parts of the magneto-optical Kerr effect~\cite{Koopmans2000,Guidoni2002}. Attributed to the laser-induced variations of the optical constants owing to the state-filling effect, it makes the determination of the genuine transient magnetization non-trivial~\cite{Razdolski2017b}. In light of these considerations, the characterization of spin filter properties of buried interfaces employing injection of hot electrons requires a novel framework wherein transient variations of magneto-optical signals can be analyzed. In particular, in the case of spin injection across an interface from ferro- into a paramagnetic material, the question of interest is pertinent to the accuracy of probing its transient magnetization with time-resolved magneto-optical techniques. It is {\it a priori} unclear to what extent the transient magneto-optical response, originating in the magnetic moment injected in the medium,
is sensitive to the distribution of the hot electrons in the reciprocal space. Reasonable evidences of the robustness of the magneto-optical response to the energy and momenta of hot electrons can turn magneto-optics into a versatile tool with ultrashort temporal resolution for the experimental investigations of (i) energy- and spin-dependent electronic transmittance of buried interfaces and (ii) transient magnetization induced by spin injection in paramagnetic media.

In this work, we analyze the ultrafast modulation of the magneto-optical properties of metals based on the injection of hot spin-polarized electrons from a ferromagnet into an adjacent metallic layer. The ballistic character of the hot electron transport is evidenced by means of the time-resolved magneto-induced second harmonic generation (mSHG), previously shown to be sensitive to ultrafast spin currents~\cite{Alekhin2017}. Using the conventional time-resolved magneto-optical Kerr technique, we demonstrate the emergence of sizable magneto-optical effects from a gold film upon the laser-induced injection of the spin current pulse from Fe on the timescale of 100 fs. The magnitude of the effect corresponds to that measured in an external magnetic field of $6.3$~T, according to the comparison of the magneto-optical coefficient obtained from our experimental results with that calculated using the results of Ref.~\cite{McGroddy1965}. In our work we introduce a framework for the analysis of the magneto-optical response of metals where the phase of the complex Kerr angle is key to the characterization of the magneto-optical properties. Applying our formalism to Au, we compare our experimental results with those obtained from the {\it ab initio} calculations as well as with other experimental and theoretical results known from the literature~\cite{McGroddy1965,Haefner1994,Etchegoin2006,Choi2014,Kimling2017}. Demonstrating striking consistency, the phase of the magneto-optical constant of Au in all relevant cases was found to be close to $0$ irrespective of the origin of magnetization, i.e. external magnetic field or spin injection. Our results thus confirm the applicability of transient magneto-optics for studying the spin injection and, subsequently, for the characterization of spin filter properties of buried metal interfaces. The formalism developed in our work is applicable to a large variety of metallic multilayers, where the interlayer spin transport can be expected to contribute to the transient variations of magneto-optical response~\cite{Rudolf2012, Eschenlohr2013, Bergeard2016, Banerjee2016, Hofherr2017}.

\section{Time-resolved magneto-optics}

To study transient magneto-optical properties of Au upon the spin current injection, we employed the time-resolved magneto-optical Kerr effect (MOKE) in the back pump-front probe experimental scheme on a Fe/Au bilayer, similar to that used in our other work~\cite{Melnikov2011}. Transient variations of the MOKE response of Au induced by the injected spins are monitored as a function of the delay between the pump and probe pulses. There are three basic magneto-optical geometries determined by mutual orientation of the magnetization $\vec{M}$, sample surface $xy$ and plane of incidence $xz$: transverse ($\vec{M}||\hat{y}$), longitudinal ($\vec{M}||\hat{x}$) and polar ($\vec{M}||\hat{z}$)~\cite{Zvezdin1997,Oppeneer2001}. In the polar and longitudinal magneto-optical geometries, the MOKE results in a change of the polarization state of the incident light. In the case of a $p$-polarized probe pulse, the MOKE leads to an appearance of the in-phase $s$-polarized wave (rotation of the polarization, $\psi_{K}^{\prime}$) and the out-of-phase $s$-polarized component (ellipticity, $\psi_{K}^{\prime\prime}$). Thus, the complete MOKE response can be represented as a complex Kerr angle $\psi_{K}$ with a phase $\varphi_{K}$:
\begin{equation}
\psi_{K}=\psi_{K}^{\prime}+i\psi_{K}^{\prime\prime}=|\psi_{K}|\cdot \exp(i\varphi_{K}).
\label{Eq_Kerr_angle}
\end{equation}

\begin{figure}[t]
\includegraphics[width=1\columnwidth]{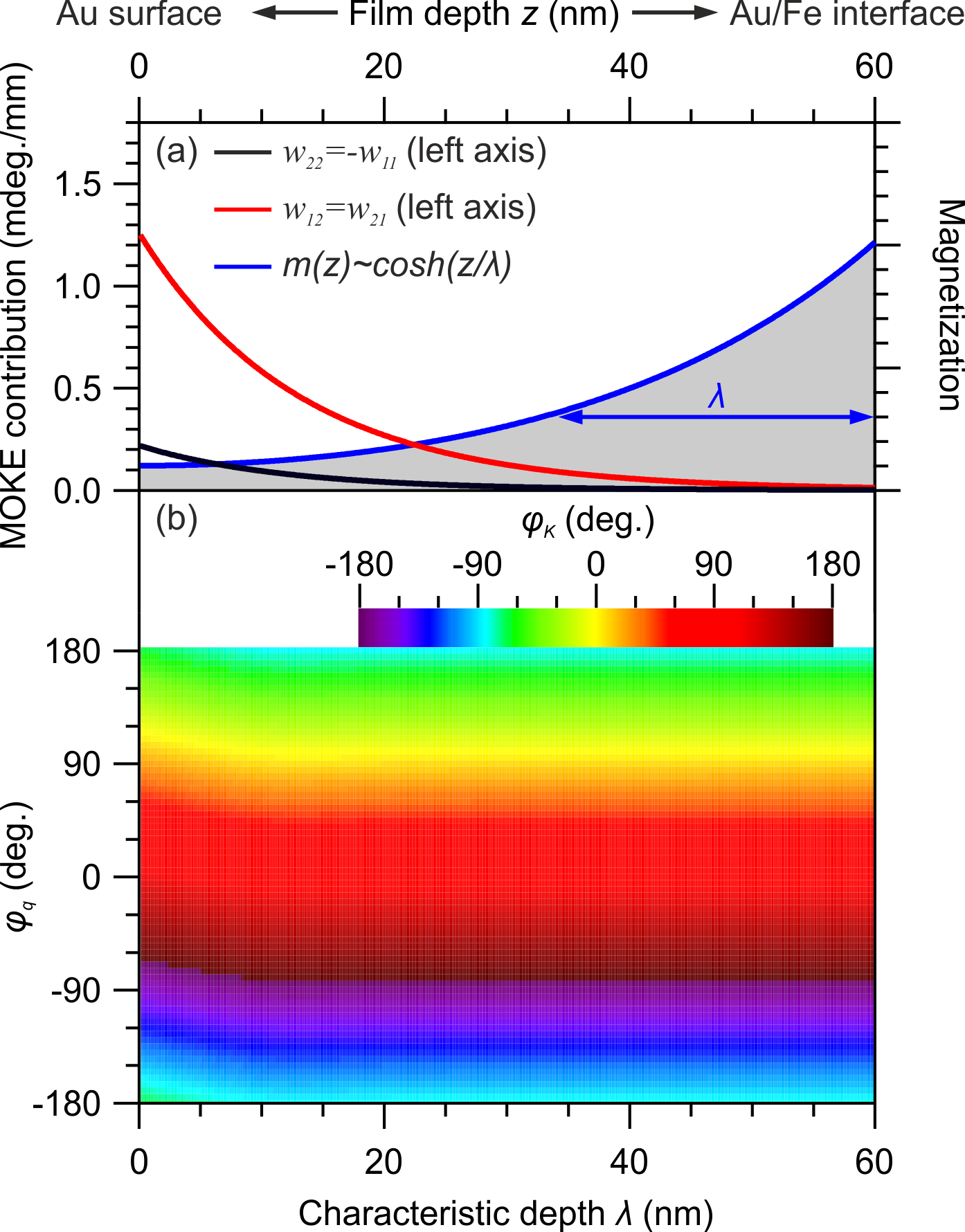}
\caption{(Color online) (a) Partial contributions $w_{ij}(z)$ of an infinitely thin magnetized Au layer to the longitudinal MOKE $\psi_K$ calculated for the photon energy $\hbar\omega=1.5$ eV, angle of incidence of the $p$-polarized light $\theta_{\rm inc}=50^{\circ}$, and $\varepsilon_{\rm Au}=-25.7+1.6i$~\cite{Johnson1972}. The dashed area illustrates a model magnetization profile $m(z)$ in a thick Au film with varied characteristic depth $\lambda$. (b) The longitudinal MOKE phase $\varphi_{K}$ calculated as a function of the magnetization profile depth $\lambda$ (see Panel a) and the phase of the complex Voigt constant of Au $\varphi_{q}$.}
\label{MOKE_sens}
\end{figure}

Within the phenomenological description of the linear magneto-optical response of an isotropic medium, the local dielectric tensor can be written in the following form~\cite{Bennemann1998}:
\begin{equation}
\label{DielectricTensor}
\hat{\varepsilon} = \varepsilon\cdot
\begin{pmatrix}
1 & iqm_z & -iqm_y\\
-iqm_z & 1 & iqm_x\\
iqm_y & -iqm_x & 1
\end{pmatrix},
\end{equation}
where
\begin{equation}
q=q^{\prime}+iq^{\prime\prime}=|q|\cdot \exp(i\varphi_{q})
\end{equation}
is a complex magneto-optical Voigt coefficient~\cite{Voigt1915}, and $\mathbf{m}=\mathbf{m}(\mathbf{r})$ is the local magnetization (which can be inhomogeneously distributed in space). Employing the 4-by-4 transfer matrix method~\cite{Zak1990}, for $\lvert qm\rvert\ll1$ one can calculate the MOKE response for any arbitrary distribution of the magnetization~\cite{Wieczorek2015}. In short, the relation between vectorial forms of the Kerr angle $\mathbf{\Psi}=(\psi_{K}^{\prime},\psi_{K}^{\prime\prime})$ and the Voigt constant $\mathbf{q}=(q^{\prime},q^{\prime\prime})$ can be written as:
\begin{equation}
\label{matrix1}
\mathbf{\Psi}=\hat{W}(\mathcal{M})\cdot\mathbf{q}.
\end{equation}
Here $\hat{W}$ is a $2\times2$ matrix with real elements $W_{22}=-W_{11}$ and $W_{12}=W_{21}$ determined by the dielectric function $\varepsilon$ and the magneto-optical geometry, and $\mathcal{M}$ illustrates the dependence of $\hat{W}$ on the distribution of the magnetic moment in the medium. The physical meaning of $\hat{W}$ is revealed by:
\begin{equation}
\label{Wmeans}
\hat{W}=\int_0^d \hat{w}(z)m(z)~dz,
\end{equation}
where the integration is performed element-wise, $m(z)$ is the spatial magnetization profile in the medium spanning from $z=0$ to $z=d$, and the components of the matrix $\hat{w}(z)$ represent the partial MOKE contributions of an infinitely thin layer positioned at a distance $z$ from the probed surface. As an illustration of the method, $w_{ij}(z)$, calculated for a Au film, are shown in Fig.~\ref{MOKE_sens},a.

In the experiments, where the MOKE rotation $\psi_{K}^{\prime}(t)$ and ellipticity $\psi_{K}^{\prime\prime}(t)$ signals are measured, it is convenient to consider their complex in-depth sensitivity function $\mathbf{w_K}=(w^{\prime}_{K},w^{\prime\prime}_{K})$:
\begin{equation}
\label{MOKE_depth_sens}
\mathbf{w_K}(z)=\hat{w}(z)\cdot\mathbf{q},
\end{equation}
so that:
\begin{equation}
\label{totalkerr}
\mathbf{\Psi}=\int_0^d\hat{w}(z)\cdot\mathbf{q}m(z)~dz=\int_0^d\mathbf{w_K}(z)m(z)~dz.
\end{equation}
In the case of a uniform magnetization profile, $m(z)\equiv M$ can be separated from the integrand, thus simplifying Eqs.~\eqref{totalkerr} to $\mathbf{\Psi}=M\cdot\int\mathbf{w_K}(z)~dz$. In time-resolved experiments, modifications of the magneto-optical response of the medium can originate either in transient variations of the magnetization profile $\Delta m(z,t)$ or in those of the magneto-optical constant $\mathbf{q}(z, t)$~\cite{Razdolski2017b}, as well as concomitant perturbations of the purely optical response $\varepsilon(z, t)$.

Note that both $\hat{W}$ and MOKE magnitude $|\psi_K|$ scale with the magnetization $M$ inside the Au layer. In the experiments where an external magnetic field induces magneto-optical response of dia- and paramagnetic media, $M$ is proportional to the field strength. At the same time, in the spin injection experiments, $M(\mathbf{r}, t)$ is {\it a priori} unknown, time-dependent, spatially inhomogeneous and depends on the injection efficiency.
For these reasons, to characterize magneto-optical properties of Au and compare our results with previous works, we consider the magnetization-independent ratio between the imaginary and real parts of the complex Voigt constant $\beta=q^{\prime\prime}/q^{\prime}$ and its phase $\varphi_{q}=\tan^{-1}\beta$. From Eq.~\eqref{matrix1} one can derive:
\begin{equation}
\label{phase}
\beta=\frac{W_{12}\cdot\psi_{K}^{\prime}-W_{11}\cdot\psi_{K}^{\prime\prime}}{W_{11}\cdot\psi_{K}^{\prime}+W_{12}\cdot\psi_{K}^{\prime\prime}}.
\end{equation}
To illustrate our approach, we have calculated the phase of the longitudinal MOKE response $\varphi_K$ in a 60 nm-thick Au film as a function of the phase of the Voigt coefficient $\varphi_q$ for a model magnetization profile $m(z)\propto\cosh(z/\lambda)$  (see Fig.~\ref{MOKE_sens}, a). This spatial profile is reminiscent of the injection of spin-polarized electrons, decaying in Au starting at the Au/Fe interface located at $z\geq60$~nm with a characteristic depth $\lambda$. The MOKE is probed from the side of the Au surface at $z=0$. The false colour plot in Fig.~\ref{MOKE_sens}, b shows how the phase of the Kerr angle $\varphi_K$ depends on the phase of the Voigt coefficient $\varphi_q$ and the characteristic depth of the magnetization profile $\lambda$. It is seen that, if $\varphi_{q}$ is considered constant as a material property, variations of $\varphi_{K}$ in the range of small $\lambda<15$~nm do not exceed $20^{\circ}$. Moreover, for larger $\lambda\gtrsim15$~nm no changes of $\varphi_{K}$ can be expected. As long as $\lambda\gtrsim15$~nm holds, transient variations of $\lambda$ upon propagation of the spin-polarized electrons in Au cannot alter the phase of $\varphi_K$. As such, we can significantly simplify the calculations by assuming that the magnetization profile $m(z,t)$ inside the probed region of the Au film is spatially homogeneous. This homogeneous, time-dependent magnetization profile $m(z,t)=M(t)$ will be used in the further analysis.

The MOKE probe is complemented here by the time-resolved mSHG technique developed in our earlier publications~\cite{Melnikov2011,Alekhin2017}. Being particularly sensitive to surfaces and interfaces of the media with inversion symmetry, this nonlinear-optical technique is more delicate in terms of data interpretation due to a complex structure of the mSHG response. The second harmonic (SH) field $E_{2\omega}=E_{e}+E_{m}$ can be represented as a sum of non-magnetic or "electronic" $E_{e}$ and magneto-induced $E_{m}$ terms which are independent of and proportional to the magnetization $M$, respectively~\cite{Pan1989}. In the dipole approximation, the SHG relies on the inversion symmetry breaking which is realized either at surfaces and interfaces or due to asymmetric distortions in the bulk, such as gradients of strain, etc. In the absence of the pump stimulus, the lack of spin polarization in Au ensures $E_{m}=0$. When the spin-polarized electrons injected across the Fe/Au interface reach the Au surface, they alter the "electronic" surface contribution $E_{e}^{\rm surf}$ and give rise to the magneto-induced field $E_{m}^{\rm surf}$, originating in the dipole polarization $P_{i}^{2\omega}=\chi_{ijk}^{(2)}E_{j}^{\omega}E_{k}^{\omega}+\chi_{ijk,l}^{(2,m)}E_{j}^{\omega}E_{k}^{\omega}M_{l}^{\rm Au}$, where $i, j, k, l = x, y, z$. It is convenient to quantify the time-dependent mSHG effects by means of the transient magnetic SH contrast $\rho_{2\omega}(t)$ arising from the interference of the $E_{e}$ and $E_{m}$ contributions:
\begin{equation}
\rho_{2\omega}(t)=\frac{I_{2\omega}^{\uparrow}(t)-I_{2\omega}^{\downarrow}(t)}{I_{2\omega}^{\uparrow}(t)+I_{2\omega}^{\downarrow}(t)}=\frac{2E_{e}E_{m}\cos\xi_{2\omega}}{|E_{e}|^{2}+|E_{m}|^2},
\end{equation}
where $I_{2\omega}^{\uparrow}(t)$ and $I_{2\omega}^{\downarrow}(t)$ are the SH intensities measured for two opposite directions of the Fe magnetization $\vec{M}_{\rm Fe}$, and $\xi_{2\omega}$ is the phase difference between $E_{e}$ and $E_{m}$. According to the symmetry properties of the mSHG at the surface of an isotropic medium or (001) face of a cubic material~\cite{Pan1989}, the non-magnetic $E_{e}$ is {\it p}-polarized in the case of {\it p}-polarized probe pulse. The magneto-induced SH component $E_{m}$ generated by the {\it p}-polarized fundamental radiation is also {\it p}-polarized in the transverse magneto-optical geometry, but {\it s}-polarized in the longitudinal one. For that reason, in order to observe the interference between $E_{m}$ and $E_{e}$, we employed the ({\it p}-in, {\it p}-out) and ({\it p}-in, $45^{\circ}$-out) polarization configurations in the transverse and longitudinal geometries, respectively.

After the laser excitation, the electrons injected into Au across the inner Fe/Au interface propagate along $\hat{z}$ towards the Au surface, thereby breaking the inversion symmetry and enabling the {\it bulk} dipole mSHG~\cite{Alekhin2017}. This flux of hot spin-polarized electrons can be characterized by a \emph{charge} current $j_{z}$ \footnote{More accurate description might treat $j_{z}$ as a \emph{particle} current since the charge will be to a large extent compensated/screened thanks to the displacement of "cold", non-spin-polarized electrons in Au.} and a \emph{spin} current (SC) $j_{z,\sigma}^{S}$ in the $\hat{z}$ direction. Here $\sigma=x,y$ indicates the orientation of the spin component governed by the magnetization in Fe $\vec{M}_{\rm Fe}$ which is aligned parallel to $\hat{y}$ and $\hat{x}$ in the transverse and longitudinal configurations, respectively. These currents produce corresponding current- and SC-induced terms in the nonlinear polarization, $\chi_{ijk}^{(2,C)}E_{j}^{\omega}E_{k}^{\omega}j_{z}$ and $\chi_{ijk}^{(2,SC)}E_{j}^{\omega}E_{k}^{\omega}j_{z,\sigma}^{S}$. These terms contribute to $E_{e}$ and $E_{m}$, as the SH fields $E_{e}^{\rm bulk}$ and $E_{m}^{\rm bulk}$~\cite{Alekhin2017}, respectively. Finally, the total non-magnetic and magneto-induced SH field component are $E_{e,m}=E_{e,m}^{\rm surf}+E_{e,m}^{\rm bulk}$, where the interface and bulk counterparts have the same symmetry and thus cannot be distinguished by employing various polarization geometries, sample orientations, etc. However, the relative magnitudes of the bulk and interface contributions can vary in different magneto-optical geometries, thereby resulting in a distinct dynamics of the transverse and longitudinal magnetic SH contrasts $\rho^T_{2\omega}(t)$ and $\rho^L_{2\omega}(t)$. Although below we will show this behavior in the Fe/Au bilayers, a detailed analysis of the mSHG response at the Au surface, including disentanglement of the bulk spin-current-induced and interface transient-magnetization-induced contributions and determination of SC pulse shape, is beyond the scope of this paper and will be given elsewhere. Retrieval of the SC pulse shape in Fe/Au/Fe trilayers was discussed earlier~\cite{Alekhin2017} along with the analysis of spin injection across the Fe/Au interface. In the following, we will focus on the analysis of the transient MOKE signals using the time-resolved mSHG data to justify the ballistic regime of the hot electron transport in Au and demonstrate how these two experimental techniques complement each other.

\section{Experimental approach}

To achieve steady excitation of SC pulses and control of their spin polarization, we utilize Au/Fe bilayers grown epitaxially on optically transparent MgO(001) substrates. Adherence to the substrate cleaning procedure and favourable film growth conditions facilitated epitaxial growth of the samples. In particular, Fe and the first nanometer of the interstitial Au layer were evaporated at $460$~K under ultrahigh vacuum. Then, the samples were cooled down and additional Au layers were evaporated at room temperature. The epitaxial growth of Fe and Au films on MgO(001) results in $[001]_{\rm Au}~||~[001]_{\rm Fe}~||~[001]_{\rm MgO}$ and $[010]_{\rm Au}~||~[110]_{\rm Fe}~||~[010]_{\rm MgO}$~\cite{Muhge1994}. The in-plane $[100]$ and $[010]$ directions in bcc-Fe correspond to the easy magnetization axes. Excellent epitaxial quality and flatness of the interfaces confirmed by transmission electron microscopy~\cite{Melnikov2011,Alekhin2017,Razdolski2017a} are essential to maintain efficient hot carrier transport within the Fe and Au layers and perform investigation of the spin filter properties of the Fe/Au interface. Another crucial advantage of the high-quality epitaxial samples is the possibility of a direct comparison of experimental results with {\it ab initio} calculations of the electron excitation in Fe~\cite{Melnikov2011} and the electronic transmittance of the Fe/Au interface~\cite{Alekhin2017}.

\begin{figure}[t]
\includegraphics[width=0.99\columnwidth]{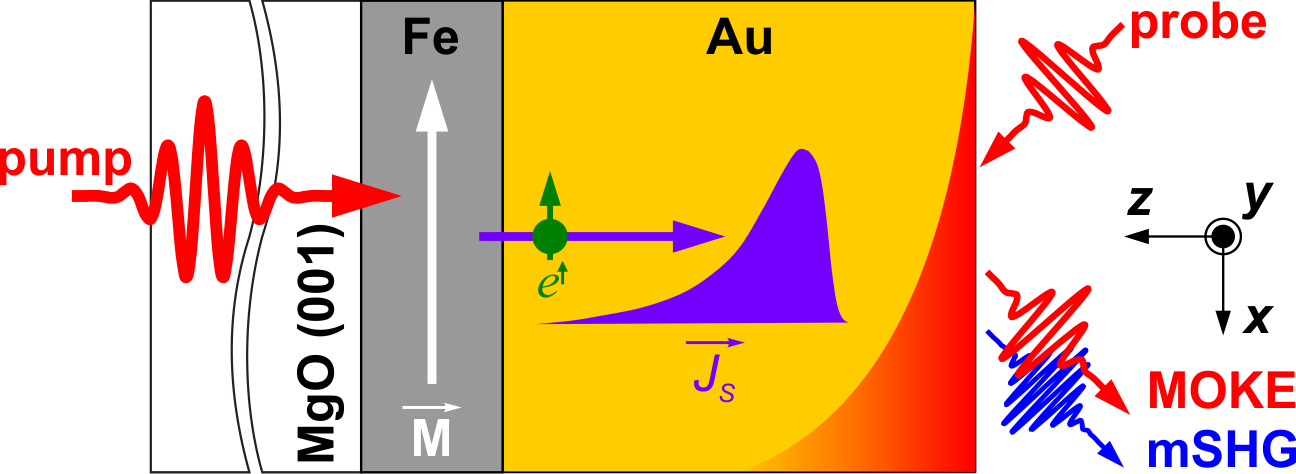}
\caption{(Color online) Experimental scheme for the injection of the laser-induced spin-polarized electrons into Au. Laser pump pulse excites hot electrons in a ferromagnetic Fe film, which propagate into the Au layer across the interface. Due to the spin- and energy-dependent transmittance of the Fe/Au interface, the injected hot electrons at elevated energies maintain sizable spin-polarization~\cite{Alekhin2017}. The induced spin polarization and spin current in Au are probed with the time-resolved MOKE and mSHG.}
\label{Exp_setup}
\end{figure}

Simultaneous time-resolved MOKE and mSHG measurements were performed using {\it p}-polarized 14-fs laser pulses (Mantis, Coherent) with the repetition rate 1 MHz and the central photon energy 1.5 eV. They were split at a power ratio 4:1 into the pump and probe beams incident at the samples at angles of $45^{\circ}$ and $50^{\circ}$ with respect to the surface normal, respectively. In a back pump-front probe scheme, pump pulses excited the Fe film, while probe pulses were used to record the magneto-optical responses from the Au side of a Fe/Au bilayer (see Fig.~\ref{Exp_setup}) as functions of the pump-probe delay. The fact that the pump and probe beams are applied from the opposite sides of the sample represents an important advantage of this configuration, as it ensures that the probed magneto-optical signals contain no spurious contributions from ferromagnetic Fe in the case of a sufficiently thick Au layer. In an alternative geometry where both pump and probe beams impinge on a thin ferromagnetic film, a more complicated analysis is required to retrieve the magneto-optical constants of a metal film from the experimental time-resolved MOKE data~\cite{Wieczorek2015,Razdolski2017b}. Furthermore, the lack of a direct optical excitation of the Au film minimizes the laser-induced variations of its optical properties, otherwise originating in the direct laser heating of the electron and lattice subsystems. As such, the only potential source of the pump-induced variations of the magneto-optical Voigt parameter $q$ and the diagonal component of the dielectric tensor $\varepsilon$ of Au is the emergence of hot spin-polarized electrons with the energy well above the Fermi level. In the following, however, we will show that these hot electrons have no significant effect on the optical properties of Au. This means that the transient MOKE signals are directly proportional to the injected magnetization in Au, which simplifies the determination of its genuine magneto-optical constants.

All samples kept at ambient conditions were mounted in such a way that the two in-plane easy axes of bcc-Fe films were aligned parallel and perpendicular to the plane of incidence. The Fe magnetization $\vec{M}_{\rm Fe}$ was set using two pairs of the Helmholtz coils producing magnetic field up to 10 mT. This field is insufficient to induce measurable magneto-optical response from Au in the absence of the pump pulse~\cite{Melnikov2011}. The mSHG output was spectrally filtered with a monochromator and detected using a photomultiplier tube. The  polarization  geometry  with  respect  to  the  SH  output was set with a polarizer in front of the monochromator. The MOKE rotation and ellipticity were measured with the two identical MOKE detectors in a balanced-photodiode scheme. In order to obtain the MOKE ellipticity, a quarter-wave plate was put in front of one of the detectors.

Owing to the in-plane magnetic anisotropy of thin ferromagnetic films, in this paper we consider the longitudinal MOKE only. The accuracy of the setup was verified by measuring the MOKE signals on a reference 32 nm-thick Fe film. The measured values of the MOKE rotation  ($\psi_{K}^{\prime}=92.1$ mdeg) and ellipticity signals ($\psi_{K}^{\prime\prime}=96.1$ mdeg)  are in an excellent agreement with the results of the 4-by-4 matrix calculation method~\cite{Zak1990} ($95.7$ and $99.8$ mdeg, respectively) using the magneto-optical constants of Fe from Ref.~\cite{Krinchik1968}. Here, the positive direction of the external magnetic field with respect to the projection of the wave-vector $\vec{k}_{i}$ of the incident light was chosen such that both MOKE rotation and ellipticity signals have the same sign as the scalar product $(\vec{M}_{\rm Fe}\cdot\vec{k}_{i})$.

\section{Experimental results}

\begin{figure}[t]
\includegraphics[width=1\columnwidth]{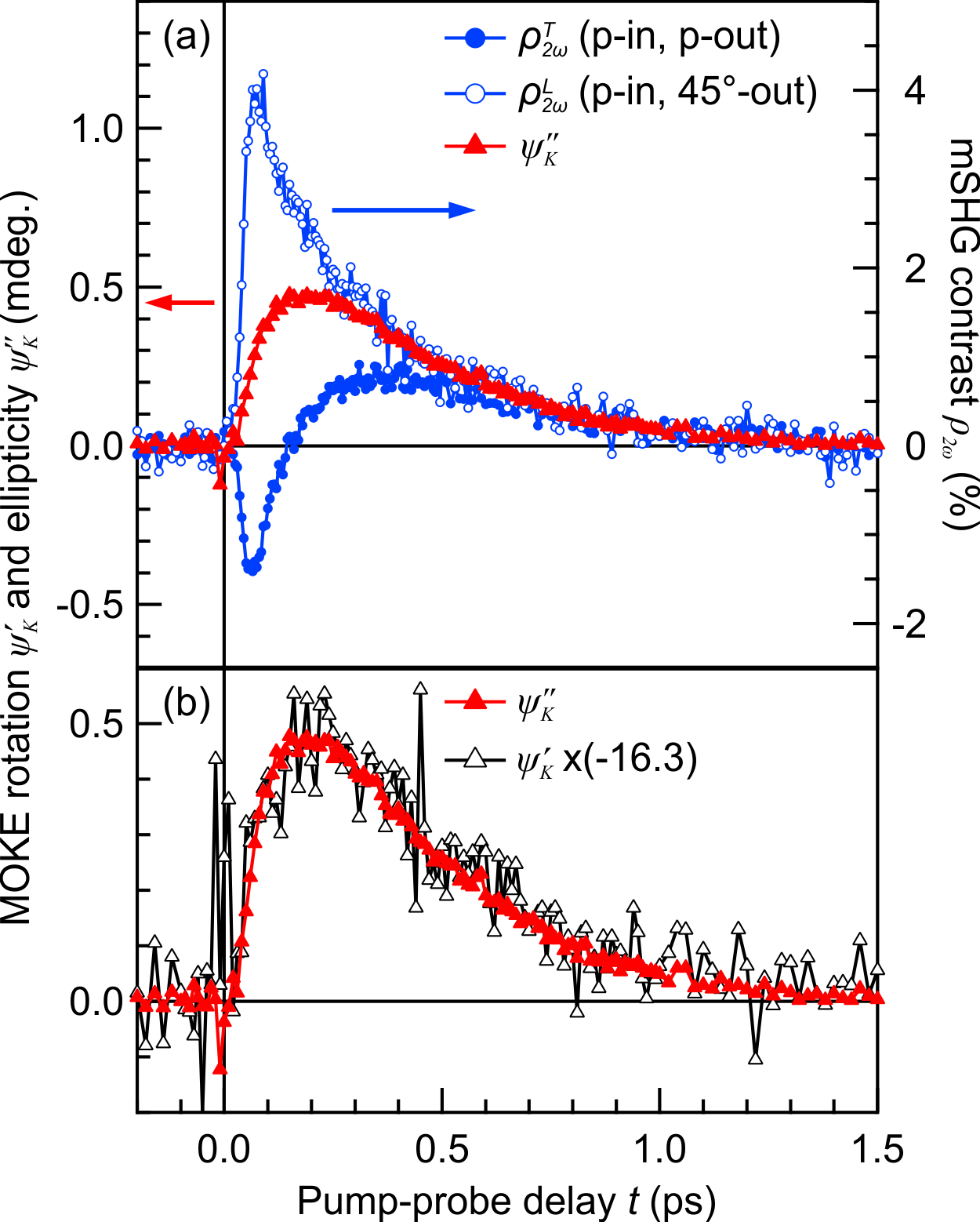}
\caption{(Color online) Transient mSHG contrasts $\rho_{2\omega}^{T}$ (blue full dots, a), $\rho_{2\omega}^{L}$ (blue open dots, a) and the MOKE ellipticity $\psi_{K}^{\prime\prime}$ (red full triangles, a and b) and rotation $\psi_{K}^{\prime}$ (black open triangles, b) measured on a 75 nm-Au/8 nm-Fe/MgO(001) sample. $\rho_{2\omega}^{T}$ was measured in the transverse magneto-optical configuration whereas the other three datasets are obtained in the longitudinal one. Note the factor -16.3 in the $\psi_{K}^{\prime}(t)$ trace.}
\label{MOKE_TL_SHG}
\end{figure}

Since our method of determining the magneto-optical constants of Au largely relies on the spin-polarized transport, we first analyze its character and efficiency in Au/Fe/MgO(001) structures with moderate Au thicknesses. Figure~\ref{MOKE_TL_SHG},a shows the transient magnetic contrasts $\rho_{2\omega}^{T,L}(t)$ of the mSHG signal and the MOKE ellipticity measured on a Fe/Au bilayer with 75 nm of Au and 8 nm of Fe in the transverse (T) and longitudinal (L) magneto-optical geometries. Both $\rho_{2\omega}^{T,L}(t)$ exhibit a sharp peak at about 60 fs after the laser excitation. However, on a larger time scale $\rho_{2\omega}^{T}(t)$ changes its sign, closely reproducing our previous results~\cite{Melnikov2011}, while $\rho_{2\omega}^{L}(t)$ remains positive. This behaviour is related to the fact that in different magneto-optical geometries different $\chi^{(2)}$ components contribute to $E_{m}$, resulting in unequal interference conditions and, consequently, distinct mSHG contrast traces $\rho_{2\omega}(t)$. Thus, the comparison of $\rho_{2\omega}^{T,L}(t)$ is vital for concluding on the sensitivity of the mSHG to both the spin polarization at the Au surface and the SC flowing in Au.

The MOKE signal is sensitive to the spin polarization only, meaning that its appearance corresponds to the presence of a non-zero magnetic moment in the probed region adjacent to the Au surface (illustrated by the red shaded area in Fig.~\ref{Exp_setup}). The width of this region is on the order of the optical penetration depth in Au ($\approx 12$ nm).
Figure~\ref{MOKE_TL_SHG},b shows transient MOKE signals measured on a Fe/Au bilayer with 75 nm of Au simultaneously with the mSHG. We note that the thickness of the Au layer is large enough to rule out magneto-optical contributions from the Fe film. It is seen that both $\psi^{\prime}_{K}(t)$, $\psi^{\prime\prime}_{K}(t)$ peak at about $t\approx 200$ fs after the pump excitation. The unipolar shape of both transient MOKE ellipticity $\psi^{\prime\prime}_{K}(t)$ and rotation $\psi^{\prime}_{K}(t)$ is in agreement with the conclusions of Ref.~\cite{Alekhin2017}: the laser-induced injection from Fe into Au is dominated by the majority electrons, while the minority ones are trapped within the Fe film. The delay between the maxima of the MOKE and mSHG transients is attributed to the fact that unlike the mSHG, the MOKE signal measures only the transient magnetization and is not sensitive to the SC leading to the appearance of sharp peaks of $\rho_{2\omega}(t)$~\footnote{Note that the peak of SC pulse retrieved in Ref.~\cite{Alekhin2017} is not as sharp as those observed in the SH magnetic contrast here. This is due to the efficient reflection of SC from the Au surface: since the contribution of reflected SC has an opposite sign, the sharp peaks in the SH magnetic contrast measured at the Au surface correspond to the steep fronts of SC pulses. A detailed discussion of this relation will be published elsewhere.}. Both magnetic mSHG contrasts $\rho_{2\omega}^{T,L}(t)$ and the MOKE rotation $\psi_{K}^{\prime}(t)$ and ellipticity $\psi_{K}^{\prime\prime}(t)$ signals vanish within the first $1-2$ picoseconds, indicating that the angular momentum of the spin-polarized electrons is either transferred to the lattice or taken away from the probed region in Au. The apparent similarity of the decay times of $\rho_{2\omega}^{T,L}(t)$ and $\psi_{K}(t)$ confirms that the sharp peaks in $\rho_{2\omega}^{T,L}(t)$ originate in the SC-induced contributions to the mSHG to which the MOKE transients are insensitive. In turn, the trailing part is determined by the spin polarization accumulated in the vicinity of the Au surface, which is probed by both magneto-optical techniques.

\begin{figure}
\includegraphics[width=1\columnwidth]{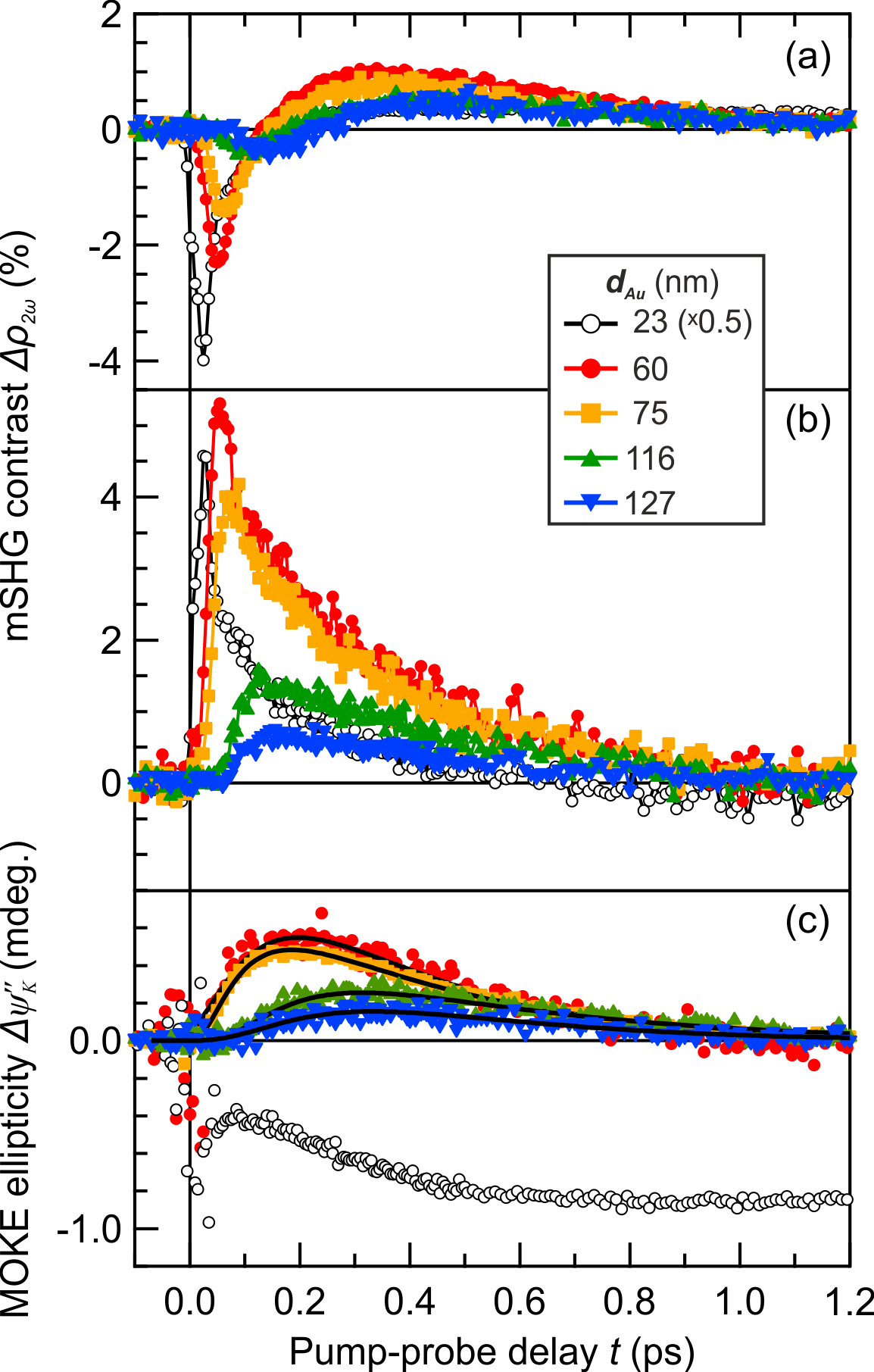}
\caption{
(Color online) Transient absolute variations of (a) transverse and (b) longitudinal mSHG contrasts $\Delta\rho_{2\omega}^{T, L}(t)$, and (c) the MOKE ellipticity $\Delta\psi_{K}^{\prime\prime}(t)$ measured on epitaxial Fe/Au bilayers with 8 nm of Fe and various thickness of the Au layer. All curves obtained for $d_{Au}=23$ nm are multiplied by a factor of 0.5 for clarity. The solid black lines in the panel (c) are the guides for the eye.}
\label{Dif_d_Au}
\end{figure}

\begin{figure}[t]
\includegraphics[width=1\columnwidth]{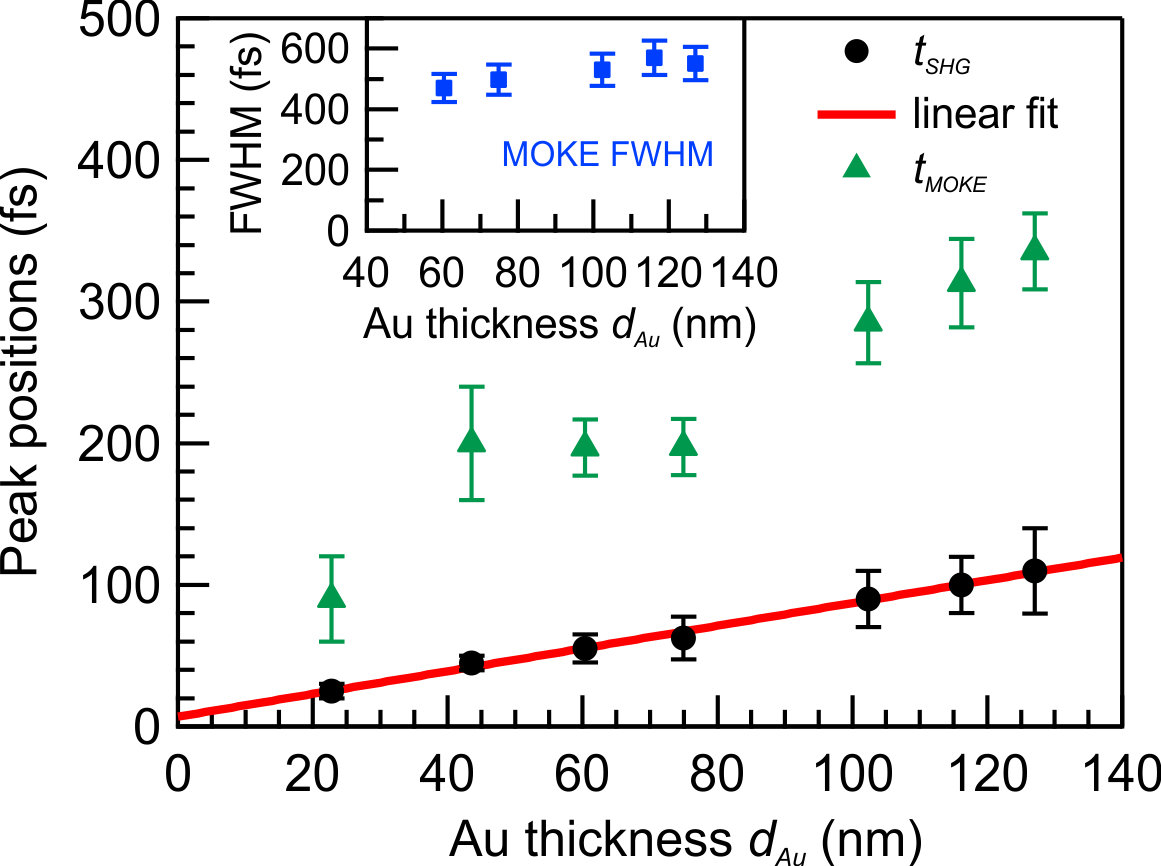}
\caption{(Color online) (a) Positions of the maximum in the transient MOKE signals (green) and the mSHG contrasts $\rho_{2\omega}^{T,L}(t)$ (black) for various Au thicknesses $d_{\rm Au}$. The solid red line is a linear fit $t=t_{0}+d_{\rm Au}/v$, indicating the ballistic character of spin transport at least for $d_{\rm Au}\leqslant130$ nm. The inset shows the temporal full width at half maximum (FWHM) of the MOKE ellipticity signal.
}
\label{Dep_d_Au}
\end{figure}

To verify further the character of the hot electron transport in Au, we measured mSHG contrasts $\rho_{2\omega}^{T,L}(t)$ and the MOKE on several samples with varied Au thickness (see Fig.~\ref{Dif_d_Au}). The negative peak in the transverse mSHG contrast $\rho_{2\omega}^{T}(t)$ gradually shallows, while its position remains proportional to the Au thickness $d_{\rm Au}$ (see Fig.~\ref{Dep_d_Au}). Similar trends can be observed in the longitudinal mSHG contrast $\rho_{2\omega}^{L}(t)$. The linear fit shown in Fig.~\ref{Dep_d_Au} is indicative of a ballistic character of the hot carrier transport, in a full agreement with our previous results~\cite{Melnikov2011}. The slope of this line gives the velocity of the hot electrons $v_{\rm Au}=(1.3\pm0.2)$ nm/fs, close to the Fermi velocity $v_{F}=1.4$~nm/fs~\cite{Brorson1987}. A small vertical offset at $d_{Au}=0$ (about $8$~fs, as obtained from the fit procedure) is below the experimental temporal resolution of 20 fs. However, the effect of the build-up of the hot electron population in Fe by virtue of carrier multiplication residing on the order of hot electron lifetimes in Fe~\cite{Zhukov2006} cannot be ruled out either.

The time-resolved MOKE traces (see Fig.~\ref{Dif_d_Au},c) show similar trends, where the maximum of the MOKE transient decreases and shifts into larger time delays for thicker Au films (see Fig.~\ref{Dep_d_Au}). Apparently, a noticeable contribution of Fe to the MOKE response for smaller Au thicknesses rules out the determination of the effective electron velocity as it was done for the SH data. For this reason, the thickness dependence of the MOKE transients requires a more intricate analysis and will be performed elsewhere. Here we merely note that the temporal full width at half maximum (FWHM) of the MOKE trace remains relatively large for all Au thicknesses (about $500$~fs, see the inset in Fig.~\ref{Dep_d_Au}). This observation is highly consistent with the electron thermalization model for the injection of hot carriers into Au~\cite{Alekhin2017} and the assumption on homogeneous distribution of the magnetization within the Au layer (see below).

To summarize our experimental observations, the transient MOKE and mSHG data confirm efficient transport of the spin-polarized electrons towards the Au surface for a wide range of Au thicknesses. The ballistic character of this transport, demonstrated for $d_{\rm Au}\leqslant130$~nm, ensures low losses of spin polarization upon propagation across the Au layer. On the other hand, our recent estimations indicate the spin density emitted across the Fe/Au interface of $7~\mu_B$/nm$^2$ under similar experimental conditions~\cite{Razdolski2017a}, thereby promising a sizable linear MOKE response of the noble metal produced by the delivered magnetic moment. For this reason, we expect that injection of hot spin-polarized electrons is capable of modifying the magneto-optical response of metal films on the timescales relevant for the ballistic electron transport. In realistic experimental conditions, i.e. for metal films with thicknesses not significantly exceeding the ballistic length ($d\leqslant100$~nm in the case of Au) the corresponding  operation rate resides on the subpicosecond timescale.

\section{\textit{Ab initio} calculations}

\begin{figure}[!h]
\includegraphics[width=0.95\columnwidth]{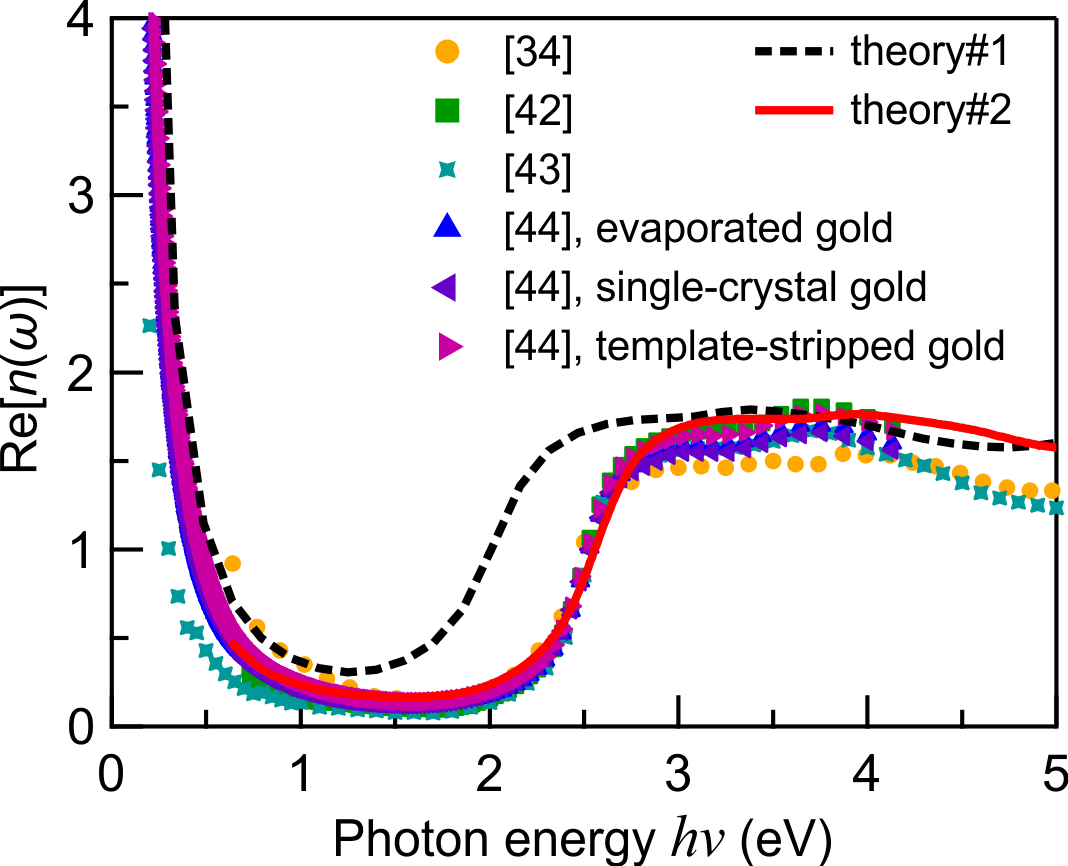}
\caption{(Color online) Comparison of measured and calculated values of the index of refraction Re~[$n(\omega )$] of gold. The experimental data are taken from Refs.~\cite{Johnson1972, mcpeak2015, Babar2015, Olmon2012}.
The theory curves have been computed without an energy shift (theory$\#$1) and with an energy shift of 0.6~eV (theory$\#$2). The used Drude parameters
were  $\sigma_{\rm D}=150\times10^{15}~{\rm s}^{-1}$, $\tau_{\rm D}=230\, {\rm Ry}^{-1}$ ($16.9$~eV$^{-1}$) for theory{\#}1 and
$\sigma_{\rm D}=240\times10^{15}~{\rm s}^{-1}$, $\tau_{\rm D}=360$ Ry$^{-1}$ ($26.5$~eV$^{-1}$) for theory{\#}2.
}
\label{fig:refl}
\end{figure}

Further insights into the non-equilibrium magneto-optical response of Au can be obtained from its comparison with the static, equilibrium case. To do this, we calculated \textit{ab initio} the longitudinal MOKE response of bulk Au in an external dc magnetic field. Thereto we first computed the optical conductivity tensor on the basis of the Kubo linear-response theory. In a single-particle formulation, suitable for \textit{ab initio} calculations, the conductivity tensor reads
\begin{eqnarray}
\label{eq:cond}
\! \! \! \sigma_{\alpha\beta}(\omega) & = & \frac{\sigma_{\rm D}^0}{1-i\omega\tau_{\rm D}}\delta_{\alpha\beta} \nonumber \\
     &-& \frac{i}{\hbar V}\! \sum_{n \neq n^\prime}\frac{f(\mathcal{E}_n)-f(\mathcal{E}_{n^\prime})}{\omega_{nn^\prime}}\frac{j_{n^{\prime}n}^{\alpha} j_{nn^\prime}^{\beta}}{\omega-\omega_{nn^\prime}+i\Gamma} .
\end{eqnarray}
Here $\hbar\omega_{nn^{\prime}}=\mathcal{E}_n-\mathcal{E}_{n^\prime}$, where $\mathcal{E}_n$ is the Bloch band energy, and $n$ represents both the band index and the Bloch wavevector $\mathbf{k}$, $f(\mathcal{E})$ is the Fermi function, $V$ is the unit cell volume, and $j^i_{nn^\prime}$ is the current operator matrix element. These quantities are calculated \textit{ab initio} within the Density Functional Theory (DFT) framework in the local density approximation (LDA) (see Ref.~\cite{Oppeneer1992} for details). The first term in Eq.~\eqref{eq:cond} is the Drude conductivity which originates from the intraband transitions ($n = n^{\prime}$) after the summation over all bands.
Although the Drude conductivity $\sigma_{\rm D}^0$ could be calculated \textit{ab initio} as well, we prefer to include the Drude term in a phenomenological way due to its strong dependence on the sample quality (e.g., scattering on defects and microstructure). The same holds true for the Drude lifetime $\tau_{\rm D}$. We assume a lifetime broadening $\Gamma = 0.01$ Ry ($0.136$~eV) and, as outlined below, two different values of the Drude conductivity and lifetime, such that the optical constant obtained in previous experiments \cite{Johnson1972,mcpeak2015,Babar2015,Olmon2012} are well described. In Fig.~\ref{fig:refl} we compare the measured and calculated refractive index of Au, Re[$n (\omega )$].
The experimental spectra are in a good overall agreement with one another. The calculated spectrum (labeled theory{\#}1) deviates from the experiments in the region of $1.5$ to $2.5$~eV. The background of this deviation is an insufficiency of the LDA approach that places the filled Au $5d$ band too high in energy, i.e., closer to the Fermi energy. A possible way of improving the spectrum is to use a scissors operation, to shift the energy bands deeper. An upward shift of $0.6$~eV of the Fermi energy leads to the curve labeled theory$\#$2, which agrees much better with the experimental data.

\begin{figure}[!h]
\includegraphics[width=0.95\columnwidth]{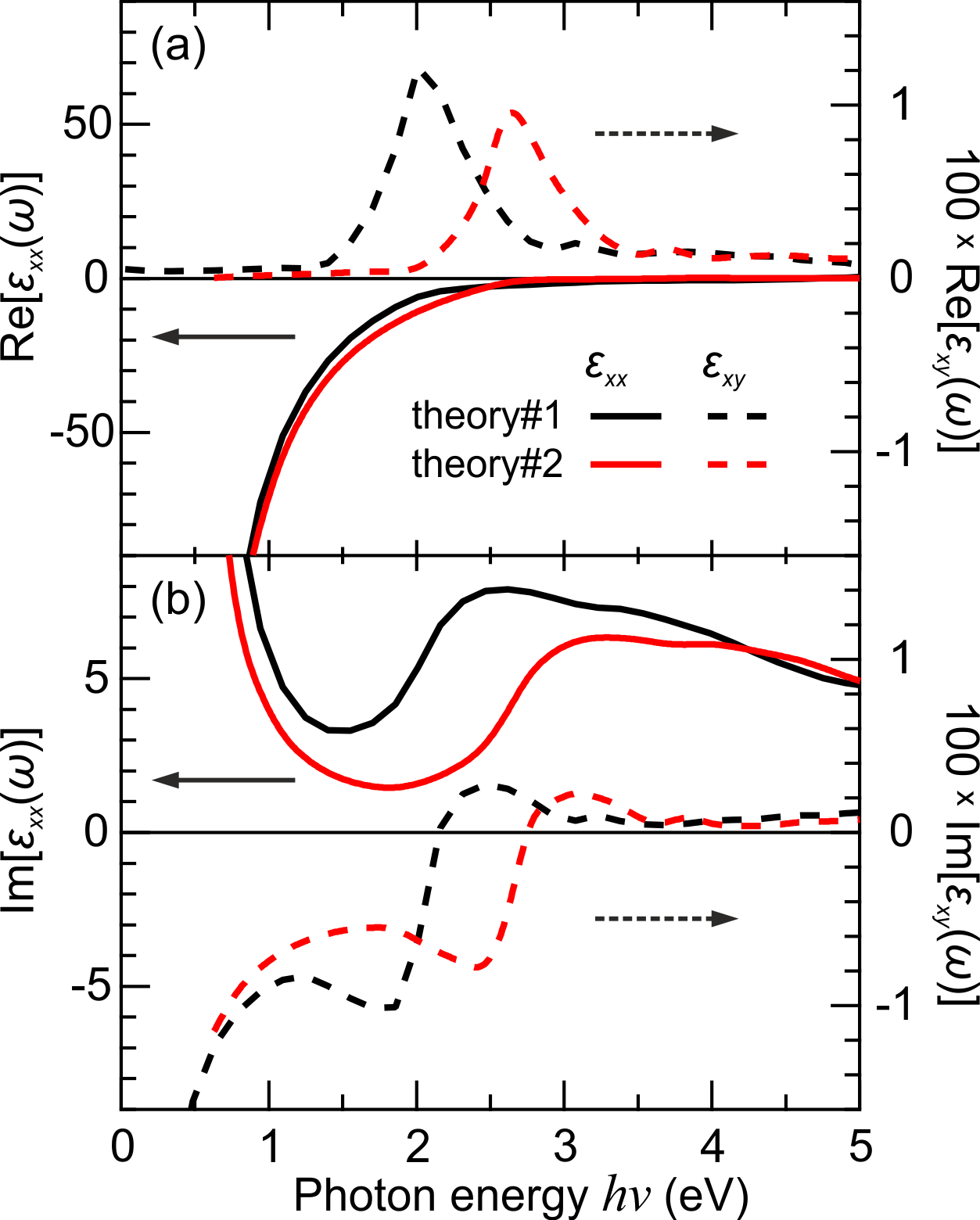}
\caption{(Color online) Calculated dielectric tensor components of Au in the presence of an applied magnetic field $H_{ext}=50$ T without energy shift (red lines) and with an energy shift $E_{\rm shift}=0.6$~eV (black lines). The Drude parameters in each case are as given in the caption of Fig.~\ref{fig:refl}.}
\label{fig:phiq}
\end{figure}

Since all the experimental data can be reformulated in terms of the dielectric tensor, we perform calculations of this tensor assuming that the Au sample is magnetized with an external magnetic field. We performed first band-structure calculations of gold when it is perturbed with a magnetic field $H_{\rm ext}=50$~T.  This magnetic field induces a magnetization of $|M|\approx10^{-3}~\mu_B$/atom. Using the resulting electronic structure of the magnetized gold we calculate the dielectric tensor for two different sets of parameters (with and without energy shift). The computed spectra are plotted in Fig.~\ref{fig:phiq}. The influence of the energy shift is clearly visible in the real and imaginary parts of the off-diagonal tensor element $\varepsilon_{xy}$. The linear scaling of the off-diagonal term of the dielectric tensor with the induced magnetization, which is in turn linear in the external magnetic field used in our calculations, has been verified up to the $500~$T field, an order of magnitude larger than those listed in Table~\ref{Tab_Models}.

The complex Kerr angle $\psi_K$ in the longitudinal geometry for $p$-polarized light can be calculated using the following equation~\cite{You1996,Oppeneer2001}:
\begin{equation}
    \psi_K=\frac{iq\bar{n}n_0}{\bar{n}^2-n_0^2}\frac{\cos\theta_{inc}\tan\theta_r}{\cos(\theta_{inc}+\theta_r)} .
\end{equation}
Here $\theta_{inc}$ and $\theta_r$ are the angles of incidence and refraction, respectively, $n_0$ is the vacuum refractive index,
and $\bar{n}$ is the averaged refractive index of Au, $\bar{n} = (n^+ + n^-)/ 2$, where $n^{\pm}$ are the refractive indices for the right and left circularly polarized eigenmodes. These can be computed from the dielectric tensor components as $(n^{\pm})^2 = \varepsilon_{xx}\pm i\varepsilon_{xy}\sin\theta_r $ \cite{Oppeneer2001}.
From Fig.~\ref{fig:refl} it is evident that the Au index of refraction computed with the scissors operation corresponds well with available experimental data. We therefore expect that the magneto-optical quantities computed under the same assumption will also tally well with our experimental data.
This is indeed the case, as shown in Table~\ref{Tab_Models} and discussed further below.
We have calculated $\psi_K\approx-0.03 + 0.31i$ at $820$~nm wavelength and $\theta_{inc}=50^{\circ}$, in a good qualitative agreement with the experimental complex Kerr angle ($\psi_K = -0.03 + 0.47i$). Computing the complex Kerr angle without the inclusion of the energy shift gives a value of $\psi_K = 0.11 + 0.36i$ (corresponding to $\varphi_q=18.5^{\circ}$) which is not able to reproduce the experimental data in a satisfactory way. With the energy shift included the MOKE ellipticity significantly exceeds the MOKE rotation in terms of absolute values, and the signs of these two quantities are opposite. Further quantitative discussion and comparison with other measurements is given below.

\section{Discussion}

It is seen in Fig.~\ref{MOKE_TL_SHG},b that the ratio between the experimental MOKE ellipticity and rotation $\psi_K^{\prime\prime}/\psi_K^{\prime}\approx-(16.3\pm0.5)$ remains constant for $t>100$ fs after the pump pulse. Moreover, this ratio was found to be almost independent of the Au thickness. These observations bring us to two important conclusions.

First, spatial magnetization profile within the probed region of the Au layer can be considered largely homogeneous at delay times $t>100$~fs. From Eqs.~\eqref{MOKE_depth_sens} and~\eqref{totalkerr}, the MOKE signals $\psi_K^{\prime}(t)$, $\psi_K^{\prime\prime}(t)$ are given by the convolution of their in-depth sensitivity functions $w_K^{\prime}(z)$, $w_K^{\prime\prime}(z)$ and the spatial magnetization profile $m(z)$. The latter is given by the transient spin polarization of hot electrons injected in Au. Taking into account the ballistic character of the electron transport and their efficient reflection at the Au surface, $m(z)$ can be reasonably well approximated by the exponential in-depth profile exemplified in Fig.~\ref{MOKE_sens}a with $\lambda=\lambda^{\rm bal}_{Au}$. As it can be inferred from Fig.~\ref{Dif_d_Au}, the ballistic length $\lambda^{\rm bal}_{Au}$ is on the order of 100~nm. On the other hand, Fig.~\ref{MOKE_sens} demonstrates that for $\lambda\gtrsim15$~nm, $\varphi_{K}$ does not change with the further increase of $\lambda$, thereby justifying $m(z)=const$ as a good approximation. The conclusion on a large value of $\lambda$ is strongly supported by the observed  $\psi_K^{\prime\prime}(t)/\psi_K^{\prime}(t)=const$ (i.e. a constant MOKE phase $\varphi_{K}$) which is insensitive to the stretching of the SC pulse with increasing Au thickness. The approximation $m(z,t)=M(t)$ is also in line with the relatively long spin emission times ($\sim 250$~fs~\cite{Alekhin2017}) as compared to the electron travel times in Au in the ballistic regime ($t= d_{\rm Au}/v\leqslant 100$ fs).

\begin{table*}[t]
\caption{\label{Tab_Models}
Dielectric tensor components and magneto-optical coefficients of Au.
}
\begin{tabular}{|l|c|c|c|c|c|c|r|r|}
\hline
\hline
Source & $\lambda$, nm  & $H_{\rm ext}$, T & $\psi_{K}$, mdeg. & $\varepsilon_{\alpha\alpha}$ & $\varepsilon_{\alpha\beta}\times10^{3}$ & $qM\times10^{4}$ & $\beta=q^{\prime\prime}/q^{\prime}$ & $\varphi_q$, deg.\\
\hline
\hspace{0.5cm}{\it Static}\\
\hline
McGroddy et al.~\cite{McGroddy1965} & 800 & $1^a$ & $0.53+0.03i$ & $-24.1+1.5i^b$  &  $-0.2-1.1i$  &  $0.46-0.05i$  & $-0.11$ & $-6.4$\\
\hline
Haefner et al.~\cite{Haefner1994} & 632.8 & 0.2 & $-^c$ & $-11.9+1.2i^d$  &  $-(0.1+14.6i)\cdot10^{-2}$  &  $0.12+0.01i$  & $0.09$ & $5.2$\\
 & &  &  &  &  $-(2.4+15.1i)\cdot10^{-2}$  &  $0.13-0.01i$  & $-0.06$ & $-3.3$\\
\hline
Drude model & $820$ & 1 & $-$ & $-34.0+1.7i^e$   &  $1.6+16.9i$  &  $-4.98+0.23i$  & $-0.05$ & $-2.6$\\
\hline
Etchegoin et al.~\cite{Etchegoin2006} & $820$ & 1 & $-$ & $-25.3+1.2i$   &  $1.23 +13.0i$  &  $-5.13 + 0.24i$  & $-0.05$ & $-2.6$\\
\hline
This work (Theory$\# 1)$ & $820$ & $25^{f}$ & $0.11+0.36i$ & $-21.5+3.2i$ & $0.8-4.6i$ & $2.1+0.7i$ & $0.33$ & 18.5\\
\hline
This work (Theory$\# 2)$ & $820$ & $50^{f}$ & $-0.03+0.31i$ & $-26.0+1.72i$ & $0.3-5.6i$ & $2.1+0.26i$ & $0.11$ & 6.8\\
\hline
\hspace{0.5cm}{\it Dynamic}\\
\hline
Hofherr et al.~\cite{Hofherr2017} & 400 & $0^g$ & $-14.8-8.3i$ & $-1.66+5.7i^b$  &  $19.4-4.9i$  &  $-32.8+14.0i$  & $-0.43$ & $-23.0$\\
\hline
Choi et al.~\cite{Choi2014} & 785 & $0^g$ & $0.085^h$ & $-22.9+1.4i^b$  &  $-0.02-0.17i$  &  $0.073-0.004i$  & $-0.06$ & $-3.4$\\
\hline
This work (exp.) & 820 & 0$^g$ & $-0.03+0.47i$ &  $-25.7+1.6i^b$  &  $0.08-7.53i$   &  $2.9 + 0.2i$   & $0.07$ & $4.2$\\
\hline
\hline
\end{tabular}
\flushleft
$^a$ The presented polar MOKE signals were normalized by the magnitude of the external magnetic field. The data is recalculated for 1 T field.\\
$^b$ From Ref.~\cite{Johnson1972}.\\
$^c$ Not applicable since transverse MOKE has been measured.\\
$^d$ From fitting the experimental data. The two values of $\varepsilon_{\alpha\beta}$ have been obtained from different experiments.\\
$^e$  Calculated using $\varepsilon_{\infty}=1$ and $\omega_p=8.89$ eV, $\tau=70.88$ meV from Ref.~\cite{Zeman1987}.\\
$^f$ In  our {\it ab initio} Augmented-Spherical-Waves framework, the estimation of the magnetic susceptibility of Au $\chi_{mol}^{th}=-3.8\times10^{-6}$~g/mol is about seven times smaller than the experimental value $\chi_{mol}^{exp}=-28.0\times10^{-6}$~g/mol~\cite{Haynes2016}.\\ 
$^g$ In the spin injection experiments, a small magnetic field ($\sim10^{-3}$ T) is used to set the magnetization of a ferromagnetic layer. This field alone is incapable of inducing sizeable MOKE signals, which is denoted here as 0.\\
$^h$ The $\psi^{\prime\prime}_K$ (MOKE ellipticity) signals were below the noise level ($5-10$ times smaller than the MOKE rotation $\psi^{\prime}_K$).\\
\end{table*}

Second, from Eq.~\eqref{matrix1} it is clear that variations of the MOKE signals could originate in the modification of either the magnetization $M$ or the magneto-optical (Voigt) coefficient $q$. The latter is responsible for the so-called optical artifact in the ultrafast MOKE transients observed in pump-probe experiments~\cite{Koopmans2000,Razdolski2017b}. These variations are attributed to the repopulation of the electronic states and concomitant modulation of the probabilities of the optical transitions. We note that despite the absence of direct laser excitation, the lack of $\Delta q$ upon spin injection in Au cannot be assumed {\it a priori}. Indeed, if the injected electrons at elevated energies occupy states which participate in the optical transitions monitored by the probe beam, the phase of $q$ can be modified. However, in this scenario the subsequent relaxation of the injected non-thermal electrons would result in the restoration of the original $q$ which could be observed as a transient variation of its phase and, consequently, the MOKE phase $\varphi_K$ as well. Thus, it is only the constant ratio between the MOKE signals that justifies the negligibility of $\Delta q$, allowing us to attribute the transient MOKE response of Au to the dynamics of its magnetization $M(t)$. As such, in this experimental configuration the magneto-optical probe delivers important information on the magnetic state of the paramagnetic metal even in the strongly non-equilibrium case, when the electronic system is perturbed on the ultrafast (femtosecond) time scale.

These conclusions allow the employment of the formalism of Eq.~(\ref{matrix1}) and use our experimental results to quantify the magneto-optical properties of Au (see Table~\ref{Tab_Models}). First of all, in its maximum the observed transient absolute MOKE signal $\lvert\psi_K\rvert$ approaches $0.5$ mdeg, about $0.5\%$ of that measured on a thick Fe film. We thus emphasize that the spin injection with a pump fluence of $10$~mJ/cm$^2$ is capable of producing similar magnitude of the Voigt vector $qM$ as the application of a magnetic field of about $6$~T in the polar MOKE experiments performed by McGroddy {\it et al.}~\cite{McGroddy1965}.
Further, the employment of inverse opto-magnetic effects represents yet another method of enabling magneto-optical activity in noble metals. We note that assuming usual one order of magnitude difference between the magnitudes of the polar and longitudinal MOKE effects, our $\varphi_K$ is comparable with that obtained at the same level of the pump fluence in the configuration of the ultrafast inverse Faraday effect~\cite{Kruglyak2005}.

Further, taking the largest transient MOKE ellipticity $\psi_{K}^{\prime\prime}(t)$ and rotation $\psi_{K}^{\prime}(t)$ signals (at around $t\approx 200$~fs), we obtain $q_{\rm Au}M_{\rm Au}=(2.9+0.2i)\cdot10^{-4}$ and, subsequently, (since $M$ is real) $\beta=q_{\rm Au}^{\prime\prime}/q_{\rm Au}^{\prime}\approx0.07$, corresponding to $\varphi_q\approx4.2^{\circ}$. As a result, we found similar values of the ratio $\beta$ and the phase $\varphi_q$ to those obtained in previous experimental works~\cite{McGroddy1965,Haefner1994} where magneto-optical activity in Au was introduced by means of an external magnetic field (denoted as {\it Static} in Table~\ref{Tab_Models}).

In what follows, we shall discuss various approaches to quantify magneto-optical response of metals. The Drude model is widely accepted as a reasonable approximation for the dielectric function of the noble metals in the spectral range far away from the resonances~\cite{Cooper1965}. In their comprehensive form, the diagonal ($\varepsilon_{\alpha\alpha}$) and off-diagonal ($\varepsilon_{\alpha\beta}$) components of the Drude dielectric function $\varepsilon_{\alpha\alpha}$ of a metal read~\cite{Bennemann1998}:
\begin{equation}
\label{Drudemodel}
\begin{split}
\varepsilon_{\alpha\alpha} & =\varepsilon_{\infty}+\frac{\omega_p^2\delta}{\omega^2\delta^2-\omega_c^2},\\
\varepsilon_{\alpha\beta} & =i\frac{\omega_p^2\omega_c}{\omega(\omega^2\delta^2-\omega_c^2)},
\end{split}
\end{equation}
where $\varepsilon_{\infty}=1$, $\omega_{p}=\sqrt{4\pi n e^2/m}$ is the plasma frequency of Au, $\omega_{c}=eB/m$ is the cyclotron frequency, $m$ and $n$ are the electron mass and concentration, respectively, and $\delta=-1-i(\omega\tau)^{-1}$ ($\tau^{-1}$ is the electron collision rate). For rare electron collisions (so that $\omega\tau\gg 1$) and small fields $B$ ($\omega_c\ll\omega$), usual $\varepsilon_{\alpha\alpha}=1-\omega_p^2/\omega^2$ and $\varepsilon_{\alpha\beta}=-i\omega_c\omega_p^2/\omega^3$ expressions can be obtained. This approach allowed Sepulveda et al.\cite{Sepulveda2010} to model the variations of the effective $\varphi_q$ of Au nanodisks in an external magnetic field in the vicinity of a surface plasmon resonance.
It should be, however, noted that various corrections to Eq.~(\ref{Drudemodel}) in the form of, for instance, $\varepsilon_{\infty}\neq 1$ have been introduced~\cite{Theye1970}, allowing Haefner et al. to obtain a better agreement with the experimental data~\cite{Haefner1994}. However, it can be shown that the introduction of a nonzero imaginary part of $\varepsilon_{\infty}$~\cite{Haefner1994} is equivalent to an additional Lorentzian resonance akin to that used for the interband transitions. Our calculations show that modifications of the real and imaginary parts of $\varepsilon_{\infty}$ result in very large variations of $\beta$ and $\varphi_q$ in the vicinity of this effective resonance. Moreover, at our wavelength of 820 nm we do not expect sizeable contributions from the interband transitions in Au. As such, we instead compare our data to the results of the uncorrected Drude model with $\varepsilon_{\infty}=1$ and other parameters from Ref.~\cite{Zeman1987}, as well as to that of the analytic Drude-based model with real $\varepsilon_{\infty}$  presented by Etchegoin {\it et al.}~\cite{Etchegoin2006} (Table~\ref{Tab_Models}). There, it is seen that both fitting the {\it static} experimental data obtained in external magnetic fields and the Drude-based approaches yield similar results.

Modern magneto-optical theories go significantly beyond the Drude approach, and the importance of spin-orbit coupling for the MOKE is already well established~\cite{Oppeneer2001}. Without going deep into the microscopic description, we recall the results of a skew scattering theory where the off-diagonal conductivity $\sigma_{\alpha\beta}$ is given by the sum of two terms \cite{Erskine1973}:

\begin{widetext}
\begin{equation}
\label{skewscattering}
\sigma_{\alpha\beta}(\omega)=\frac{\omega_p^2}{4\pi}\frac{\Omega}{\Omega^2+(1/\tau+i\omega)^2}-\frac{\omega_p^2}{4\pi}\frac{n^{\uparrow}-n^{\downarrow}}{n^{\uparrow}+n^{\downarrow}}\frac{P_0}{ev_F}\left(1-\frac{i\omega(1/\tau+i\omega)}{\Omega^2+(1/\tau+i\omega)^2}\right)
\end{equation}
\end{widetext}

Here, $P_0$ is a maximum macroscopic dipole moment, $n^{\uparrow}$, $n^{\downarrow}$ are the numbers of spin-up and spin-down electrons per unit volume, $\Omega$ is an effective frequency accounting for the spin-orbit asymmetric scattering in metals with non-zero magnetization~\cite{Fivaz1969}.
We note that $\Omega\ll 1/\tau\ll\omega$ is usually assumed (the high-frequency limit), so that the first term in Eq.~\eqref{skewscattering} acquires the familiar Drude-like form (cf.~Eq.~\ref{Drudemodel} and discussion thereafter) of $-(\omega_p/\omega)^2\cdot\Omega/4\pi$ (recall the $\varepsilon_{\alpha\beta}=\delta_{\alpha\beta}+4\pi i\sigma_{\alpha\beta}/\omega$ relation) if $\Omega$ is attributed to the cyclotrone frequency $\omega_c$.
In static experiments on para- and diamagnetic metals both $\Omega$ and $n^{\uparrow}-n^{\downarrow}$ are proportional to the external magnetic field $B$ so that the phase of complex $\sigma_{\alpha\beta}$ is independent of $B$. Simplifying Eq.~\eqref{skewscattering} in the high-frequency limit, for the phase of the off-diagonal conductivity we get $\lvert\tan\varphi_{\sigma}\rvert\sim\omega\tau\gg 1$. Recalling that $\varepsilon_{\alpha\beta}=iq\varepsilon_{\alpha\alpha}$ and neglecting the imaginary part of $\varepsilon_{\alpha\alpha}$ in noble metals far away from the interband transition resonances, the phase of the magneto-optical coefficient can be estimated as $\varphi_{q}\sim(\omega\tau)^{-1}\ll 1$.

Remarkably, this estimation is in agreement with the results of both {\it static} and {\it dynamic} experiments (Table~\ref{Tab_Models}), among which we outline the works of Choi et al.~\cite{Choi2014} and Kimling et al.~\cite{Kimling2017}. Importantly, relatively long ($0.8$~ps) pump pulses responsible for the electronic injection ensure effective electronic thermalization during the optical excitation, so that no electrons at elevated energies ($\sim 1$ eV) populate the probed Au region. As such, in those experiments the magneto-optical response was introduced by already {\it thermalized} electrons. There, $\hbar\Omega$ was attributed to the strength of the spin-orbit coupling ($\sim0.1-1$~eV in Au), still resulting in $\Omega\ll\omega_p$ and, consequently, $\varphi_q\approx0$. Note that a strongly non-zero $\varphi_q$ found by Hofherr {\it et al}.~\cite{Hofherr2017} was registered at a probe wavelength of $\lambda=400$~nm , enabling the contribution from the interband transitions in Au. More importantly, the front pump-front probe Complex-MOKE experimental geometry precludes the unambiguous determination of $q_{\rm Au}$ from the experimentally observed MOKE signals $\psi_{K}^{\prime}$, $\psi_{K}^{\prime\prime}$~\footnote{Indeed, if the detection axis in the Ni/Au bilayer on the complex plain is set perpendicular to the MOKE vector of Ni (termed by the authors "orthogonal data set"), its non-zero angle with the MOKE vector of Au results in unequal corrections of the observed $\psi_{K}^{\prime}$, $\psi_{K}^{\prime\prime}$ thereby modifying $\beta$. Moreover, the polarization state of light reflected from Au will undergo additional changes upon propagating back through the Ni layer, meaning that
the corresponding $\beta$ value shown in Table~\ref{Tab_Models} for Ref.~\cite{Hofherr2017} can only be treated as an estimate.}. Finally, Elezzabi et al.~\cite{Elezzabi1996} do not discuss the imaginary part of the transient MOKE response $\psi_{K}^{\prime\prime}$, thereby ruling out the determination of $\varphi_q$.

These results demonstrate that far away from the resonances attributed to, e.g., interband transitions, the magneto-optical response of noble metals (such as Au) on the ultrafast timescale is determined mostly by the transient spin polarization and not by the optical state-filling effects which are often concomitant with the direct laser excitation. This observation is consistent with the fact that our calculations of the static MOKE and optical spectra are in agreement with the measurements presented in Table~\ref{Tab_Models}. The {\it ab initio} calculations labeled Theory\#2 compare sufficiently well with our experimental results and available data from the literature. To perform the MOKE calculation, we assumed an equilibrium magnetization as induced by a static magnetic field. Applying an external magnetic field $H_{ext}=50$ T in a self-consistent calculation of the electronic structure, we obtained a static magnetization of $|M|\approx10^{-3}~\mu_B$ per Au atom. The magneto-optical Voigt vector $qM$, calculated for such electronic structure, is in a good agreement with our experimental results and with the MOKE measurements in a static magnetic field, performed by McGroddy {\it et al.}~\cite{McGroddy1965}. However, given that the magneto-optical coefficients depend linearly on the magnetic field, comparison of the magnitudes of $qM$, found using our experimental results and those from Ref.~\cite{McGroddy1965}, provides the actual value of the corresponding magnetic field of $6.3$ T (see Table~\ref{Tab_Models}), while the applied magnetic field used in our calculations is much larger. Using the reported magnetic susceptibility of Au~\cite{Haynes2016}, we observe that a magnetic field $H_{ext}=6.3$ T implies a magnetization of the material $|M|\approx10^{-3}~\mu_B$/atom, i.e.,  the same as the one we calculated with an applied magnetic field $B=50$ T in the band structure method. Using our {\it ab initio} Augmented-Spherical-Waves framework to estimate the magnetic susceptibility, we obtain $\chi_{mol}^{th}=-3.8\times10^{-6}$~g/mol which is about 7 times smaller than the experimental value $\chi_{mol}^{exp}=-28.0\times10^{-6}$~g/mol. This difference could occur due to the fact that we compute only the spin susceptibility and neglect the orbital susceptibility.

The striking consistency between various measurements and calculations of the phase of the magneto-optical coefficient $\varphi_q$ is very unlikely to be a mere coincidence. In contrast, we argue that this $\varphi_q$ is an inherent material property which can be monitored by the transient MOKE in the spin injection experiments. Indeed, it is known that in non-dissipative media the diagonal and non-diagonal elements of the dielectric tensor~\eqref{DielectricTensor} are purely real and imaginary, respectively~\cite{Pershan1963}, meaning that $q$ is real, i.e. $\varphi_q=0$. Taking into account low absorption at 1.5~eV in Au (well below the interband transitions), this result is highly consistent with our observations. Interestingly, although this reasoning is usually applied to optically transparent media, our analysis expands it onto the non-transparent metals where the small absorption is limited by the electron collision rate, $\omega\tau\gg 1$, resulting in $|\varepsilon^{\prime\prime}_{\alpha \alpha }| \ll |\varepsilon^{\prime}_{\alpha \alpha }|$ and  $|q^{\prime\prime}| \ll |q^{\prime}|$. The presence of hot electrons in the vicinity of the Au surface reduces the optical reflectivity by less than 1\% and therefore does not change the above relations. In light of Eq.~\eqref{eq:cond}, it is also hard to expect significant variations of $|q|$, which allows to attribute the observed MOKE response solely to the build-up of spin polarization in Au. Thus, from the point of view of inducing spin polarization, the injection of hot electrons is equivalent to the application of the external magnetic field. However, it operates on much faster timescales governed by the rates of electron thermalization and transport, well below $1$~ps.

The injection is efficient only when the high energy electrons in an adjacent ferromagnetic layer exist, i.e. before the equilibration of the electronic subsystem~\cite{Alekhin2017}. As it has been shown for the Fe/Au bilayer, once the laser-excited electrons in Fe are thermalized, the spin injection stops, meaning that the interface ceases to function as an efficient spin filter. As such, laser-induced spin injection can be envisioned as a powerful tool for the investigation of the spin properties of buried interfaces. Owing to the lack of modification of the optical constants upon laser-induced spin injection, in such time-resolved MOKE experiments the problem of distinguishing magnetism and optics \cite{Koopmans2000,Razdolski2017b} is suppressed. In light of these considerations, further experimental studies of the spin injection into para- and diamagnetic metals across various buried interfaces, as well as a direct comparison of the spin-injection-induced magneto-optical activity with and without strong external magnetic fields remain an attractive perspective.

\section{Conclusions}

To summarize, we have demonstrated spin injection-induced transient MOKE response from Au on the subpicosecond timescale. Complementary time-resolved mSHG experiments on a series of Fe/Au bilayers corroborate the ballistic character of the hot spin-polarized electrons transport towards the Au surface at least for $d_{\rm Au}\leqslant130$ nm. Employing the transfer matrix approach for the analysis of the transient MOKE phase, we have quantified the magneto-optical response of Au. The results are discussed within the framework of previous experimental findings and compared to those given by the Drude model and skew scattering theory, as well as obtained from the {\it ab initio} calculations with strong dc magnetic fields. Our analysis shows that the MOKE phase remains highly consistent throughout various measurements, indicating that in the spin injection experiments, the MOKE transients truly reflect the dynamics of spin polarization in the probed region. We further outline rich perspectives of our method in application to the studies of the electron thermalization dynamics and spin filter properties of buried interfaces on the subpicosecond timescale.

\begin{acknowledgments}
The authors thank M.~Wolf for support, P.~Maldonado for stimulating discussions. Funding by the Deutsche Forschungsgemeinschaft through ME 3570/1, SFB 616, AL2143/2-1 and SFB-TRR 227 as well as by the EU 7-th framework program through CRONOS (grant No. 280879). M.B.\ and P.M.O.\ acknowledge funding from the Swedish Research Council (VR), the K.\ and A.\ Wallenberg Foundation (Grant No.\ 2015.0060), the European Union's Horizon2020 Research and Innovation Programme under Grant agreement No.\ 737709 (FEMTOTERABYTE), and the Swedish National Infrastructure for Computing (SNIC) for computer time.
 \end{acknowledgments}

\bibliography{Bibliography}

\begin{thebibliography}{55}
\expandafter\ifx\csname natexlab\endcsname\relax\def\natexlab#1{#1}\fi
\expandafter\ifx\csname bibnamefont\endcsname\relax
  \def\bibnamefont#1{#1}\fi
\expandafter\ifx\csname bibfnamefont\endcsname\relax
  \def\bibfnamefont#1{#1}\fi
\expandafter\ifx\csname citenamefont\endcsname\relax
  \def\citenamefont#1{#1}\fi
\expandafter\ifx\csname url\endcsname\relax
  \def\url#1{\texttt{#1}}\fi
\expandafter\ifx\csname urlprefix\endcsname\relax\def\urlprefix{URL }\fi
\providecommand{\bibinfo}[2]{#2}
\providecommand{\eprint}[2][]{\url{#2}}

\bibitem[{\citenamefont{Kirilyuk et~al.}(2010)\citenamefont{Kirilyuk, Kimel,
  and Rasing}}]{Kirilyuk2010}
\bibinfo{author}{\bibfnamefont{A.}~\bibnamefont{Kirilyuk}},
  \bibinfo{author}{\bibfnamefont{A.~V.} \bibnamefont{Kimel}}, \bibnamefont{and}
  \bibinfo{author}{\bibfnamefont{T.}~\bibnamefont{Rasing}},
  \bibinfo{journal}{Rev. Mod. Phys.} \textbf{\bibinfo{volume}{82}},
  \bibinfo{pages}{2731} (\bibinfo{year}{2010}), ISSN \bibinfo{issn}{0034-6861}.

\bibitem[{\citenamefont{Kirilyuk et~al.}(2013)\citenamefont{Kirilyuk, Kimel,
  and Rasing}}]{Kirilyuk2013}
\bibinfo{author}{\bibfnamefont{A.}~\bibnamefont{Kirilyuk}},
  \bibinfo{author}{\bibfnamefont{A.~V.} \bibnamefont{Kimel}}, \bibnamefont{and}
  \bibinfo{author}{\bibfnamefont{T.}~\bibnamefont{Rasing}},
  \bibinfo{journal}{Rep. Prog. Phys.} \textbf{\bibinfo{volume}{76}},
  \bibinfo{pages}{026501} (\bibinfo{year}{2013}).

\bibitem[{\citenamefont{Beaurepaire et~al.}(1996)\citenamefont{Beaurepaire,
  Merle, Daunois, and Bigot}}]{Beaurepaire1996}
\bibinfo{author}{\bibfnamefont{E.}~\bibnamefont{Beaurepaire}},
  \bibinfo{author}{\bibfnamefont{J.}~\bibnamefont{Merle}},
  \bibinfo{author}{\bibfnamefont{A.}~\bibnamefont{Daunois}}, \bibnamefont{and}
  \bibinfo{author}{\bibfnamefont{J.}~\bibnamefont{Bigot}},
  \bibinfo{journal}{Phys. Rev. Lett.} \textbf{\bibinfo{volume}{76}},
  \bibinfo{pages}{4250} (\bibinfo{year}{1996}), ISSN \bibinfo{issn}{1079-7114}.

\bibitem[{\citenamefont{Hohlfeld et~al.}(2000)\citenamefont{Hohlfeld,
  Wellershoff, G{\"{u}}dde, Conrad, J{\"{a}}hnke, and Matthias}}]{Hohlfeld2000}
\bibinfo{author}{\bibfnamefont{J.}~\bibnamefont{Hohlfeld}},
  \bibinfo{author}{\bibfnamefont{S.~S.} \bibnamefont{Wellershoff}},
  \bibinfo{author}{\bibfnamefont{J.}~\bibnamefont{G{\"{u}}dde}},
  \bibinfo{author}{\bibfnamefont{U.}~\bibnamefont{Conrad}},
  \bibinfo{author}{\bibfnamefont{V.}~\bibnamefont{J{\"{a}}hnke}},
  \bibnamefont{and} \bibinfo{author}{\bibfnamefont{E.}~\bibnamefont{Matthias}},
  \bibinfo{journal}{Chem. Phys.} \textbf{\bibinfo{volume}{251}},
  \bibinfo{pages}{237} (\bibinfo{year}{2000}), ISSN \bibinfo{issn}{03010104}.

\bibitem[{\citenamefont{Walowski and M{\"{u}}nzenberg}(2016)}]{Walowski2016}
\bibinfo{author}{\bibfnamefont{J.}~\bibnamefont{Walowski}} \bibnamefont{and}
  \bibinfo{author}{\bibfnamefont{M.}~\bibnamefont{M{\"{u}}nzenberg}},
  \bibinfo{journal}{J. Appl. Phys.} \textbf{\bibinfo{volume}{120}},
  \bibinfo{pages}{140901} (\bibinfo{year}{2016}).

\bibitem[{\citenamefont{Razdolski
  et~al.}(2017{\natexlab{a}})\citenamefont{Razdolski, Alekhin, Martens,
  B{\"{u}}rstel, Diesing, M{\"{u}}nzenberg, Bovensiepen, and
  Melnikov}}]{Razdolski2017b}
\bibinfo{author}{\bibfnamefont{I.}~\bibnamefont{Razdolski}},
  \bibinfo{author}{\bibfnamefont{A.}~\bibnamefont{Alekhin}},
  \bibinfo{author}{\bibfnamefont{U.}~\bibnamefont{Martens}},
  \bibinfo{author}{\bibfnamefont{D.}~\bibnamefont{B{\"{u}}rstel}},
  \bibinfo{author}{\bibfnamefont{D.}~\bibnamefont{Diesing}},
  \bibinfo{author}{\bibfnamefont{M.}~\bibnamefont{M{\"{u}}nzenberg}},
  \bibinfo{author}{\bibfnamefont{U.}~\bibnamefont{Bovensiepen}},
  \bibnamefont{and} \bibinfo{author}{\bibfnamefont{A.}~\bibnamefont{Melnikov}},
  \bibinfo{journal}{J. Phys.: Condens. Matter} \textbf{\bibinfo{volume}{29}},
  \bibinfo{pages}{174002} (\bibinfo{year}{2017}{\natexlab{a}}), ISSN
  \bibinfo{issn}{0953-8984}.

\bibitem[{\citenamefont{Wieczorek et~al.}(2015)\citenamefont{Wieczorek,
  Eschenlohr, Weidtmann, R{\"{o}}sner, Bergeard, Tarasevitch, Wehling, and
  Bovensiepen}}]{Wieczorek2015}
\bibinfo{author}{\bibfnamefont{J.}~\bibnamefont{Wieczorek}},
  \bibinfo{author}{\bibfnamefont{A.}~\bibnamefont{Eschenlohr}},
  \bibinfo{author}{\bibfnamefont{B.}~\bibnamefont{Weidtmann}},
  \bibinfo{author}{\bibfnamefont{M.}~\bibnamefont{R{\"{o}}sner}},
  \bibinfo{author}{\bibfnamefont{N.}~\bibnamefont{Bergeard}},
  \bibinfo{author}{\bibfnamefont{A.}~\bibnamefont{Tarasevitch}},
  \bibinfo{author}{\bibfnamefont{T.~O.} \bibnamefont{Wehling}},
  \bibnamefont{and}
  \bibinfo{author}{\bibfnamefont{U.}~\bibnamefont{Bovensiepen}},
  \bibinfo{journal}{Phys. Rev. B - Condensed Matter and Materials Physics}
  \textbf{\bibinfo{volume}{92}}, \bibinfo{pages}{174410}
  (\bibinfo{year}{2015}), ISSN \bibinfo{issn}{1550235X}.

\bibitem[{\citenamefont{Battiato et~al.}(2010)\citenamefont{Battiato, Carva,
  and Oppeneer}}]{Battiato2010}
\bibinfo{author}{\bibfnamefont{M.}~\bibnamefont{Battiato}},
  \bibinfo{author}{\bibfnamefont{K.}~\bibnamefont{Carva}}, \bibnamefont{and}
  \bibinfo{author}{\bibfnamefont{P.~M.} \bibnamefont{Oppeneer}},
  \bibinfo{journal}{Phys. Rev. Lett.} \textbf{\bibinfo{volume}{105}},
  \bibinfo{pages}{27203} (\bibinfo{year}{2010}).

\bibitem[{\citenamefont{Malinowski et~al.}(2008)\citenamefont{Malinowski,
  {Dalla Longa}, Rietjens, Paluskar, Huijink, Swagten, and
  Koopmans}}]{Malinowski2008}
\bibinfo{author}{\bibfnamefont{G.}~\bibnamefont{Malinowski}},
  \bibinfo{author}{\bibfnamefont{F.}~\bibnamefont{{Dalla Longa}}},
  \bibinfo{author}{\bibfnamefont{J.~H.~H.} \bibnamefont{Rietjens}},
  \bibinfo{author}{\bibfnamefont{P.~V.} \bibnamefont{Paluskar}},
  \bibinfo{author}{\bibfnamefont{R.}~\bibnamefont{Huijink}},
  \bibinfo{author}{\bibfnamefont{H.~J.~M.} \bibnamefont{Swagten}},
  \bibnamefont{and} \bibinfo{author}{\bibfnamefont{B.}~\bibnamefont{Koopmans}},
  \bibinfo{journal}{Nat. Phys.} \textbf{\bibinfo{volume}{4}},
  \bibinfo{pages}{855} (\bibinfo{year}{2008}), ISSN \bibinfo{issn}{1745-2473}.

\bibitem[{\citenamefont{Melnikov et~al.}(2011)\citenamefont{Melnikov,
  Razdolski, Wehling, Papaioannou, Roddatis, Fumagalli, Aktsipetrov,
  Lichtenstein, and Bovensiepen}}]{Melnikov2011}
\bibinfo{author}{\bibfnamefont{A.}~\bibnamefont{Melnikov}},
  \bibinfo{author}{\bibfnamefont{I.}~\bibnamefont{Razdolski}},
  \bibinfo{author}{\bibfnamefont{T.~O.} \bibnamefont{Wehling}},
  \bibinfo{author}{\bibfnamefont{E.~T.} \bibnamefont{Papaioannou}},
  \bibinfo{author}{\bibfnamefont{V.}~\bibnamefont{Roddatis}},
  \bibinfo{author}{\bibfnamefont{P.}~\bibnamefont{Fumagalli}},
  \bibinfo{author}{\bibfnamefont{O.}~\bibnamefont{Aktsipetrov}},
  \bibinfo{author}{\bibfnamefont{A.~I.} \bibnamefont{Lichtenstein}},
  \bibnamefont{and}
  \bibinfo{author}{\bibfnamefont{U.}~\bibnamefont{Bovensiepen}},
  \bibinfo{journal}{Phys. Rev. Lett.} \textbf{\bibinfo{volume}{107}},
  \bibinfo{pages}{76601} (\bibinfo{year}{2011}), ISSN \bibinfo{issn}{00319007},
  \eprint{1103.5310}.

\bibitem[{\citenamefont{Eschenlohr et~al.}(2013)\citenamefont{Eschenlohr,
  Battiato, Maldonado, Pontius, Kachel, Holldack, Mitzner, F{\"{o}}hlisch,
  Oppeneer, and Stamm}}]{Eschenlohr2013}
\bibinfo{author}{\bibfnamefont{A.}~\bibnamefont{Eschenlohr}},
  \bibinfo{author}{\bibfnamefont{M.}~\bibnamefont{Battiato}},
  \bibinfo{author}{\bibfnamefont{P.}~\bibnamefont{Maldonado}},
  \bibinfo{author}{\bibfnamefont{N.}~\bibnamefont{Pontius}},
  \bibinfo{author}{\bibfnamefont{T.}~\bibnamefont{Kachel}},
  \bibinfo{author}{\bibfnamefont{K.}~\bibnamefont{Holldack}},
  \bibinfo{author}{\bibfnamefont{R.}~\bibnamefont{Mitzner}},
  \bibinfo{author}{\bibfnamefont{A.}~\bibnamefont{F{\"{o}}hlisch}},
  \bibinfo{author}{\bibfnamefont{P.~M.} \bibnamefont{Oppeneer}},
  \bibnamefont{and} \bibinfo{author}{\bibfnamefont{C.}~\bibnamefont{Stamm}},
  \bibinfo{journal}{Nat. Mater.} \textbf{\bibinfo{volume}{12}},
  \bibinfo{pages}{332} (\bibinfo{year}{2013}), ISSN \bibinfo{issn}{14761122}.

\bibitem[{\citenamefont{Turgut et~al.}(2013)\citenamefont{Turgut,
  La-O-Vorakiat, Shaw, Grychtol, Nembach, Rudolf, Adam, Aeschlimann, Schneider,
  Silva et~al.}}]{Turgut2013}
\bibinfo{author}{\bibfnamefont{E.}~\bibnamefont{Turgut}},
  \bibinfo{author}{\bibfnamefont{C.}~\bibnamefont{La-O-Vorakiat}},
  \bibinfo{author}{\bibfnamefont{J.~M.} \bibnamefont{Shaw}},
  \bibinfo{author}{\bibfnamefont{P.}~\bibnamefont{Grychtol}},
  \bibinfo{author}{\bibfnamefont{H.~T.} \bibnamefont{Nembach}},
  \bibinfo{author}{\bibfnamefont{D.}~\bibnamefont{Rudolf}},
  \bibinfo{author}{\bibfnamefont{R.}~\bibnamefont{Adam}},
  \bibinfo{author}{\bibfnamefont{M.}~\bibnamefont{Aeschlimann}},
  \bibinfo{author}{\bibfnamefont{C.~M.} \bibnamefont{Schneider}},
  \bibinfo{author}{\bibfnamefont{T.~J.} \bibnamefont{Silva}},
  \bibnamefont{et~al.}, \bibinfo{journal}{Phys. Rev. Lett.}
  \textbf{\bibinfo{volume}{110}}, \bibinfo{pages}{197201}
  (\bibinfo{year}{2013}), ISSN \bibinfo{issn}{00319007}.

\bibitem[{\citenamefont{Alekhin et~al.}(2017)\citenamefont{Alekhin, Razdolski,
  Ilin, Meyburg, Diesing, Roddatis, Rungger, Stamenova, Sanvito, Bovensiepen
  et~al.}}]{Alekhin2017}
\bibinfo{author}{\bibfnamefont{A.}~\bibnamefont{Alekhin}},
  \bibinfo{author}{\bibfnamefont{I.}~\bibnamefont{Razdolski}},
  \bibinfo{author}{\bibfnamefont{N.}~\bibnamefont{Ilin}},
  \bibinfo{author}{\bibfnamefont{J.~P.} \bibnamefont{Meyburg}},
  \bibinfo{author}{\bibfnamefont{D.}~\bibnamefont{Diesing}},
  \bibinfo{author}{\bibfnamefont{V.}~\bibnamefont{Roddatis}},
  \bibinfo{author}{\bibfnamefont{I.}~\bibnamefont{Rungger}},
  \bibinfo{author}{\bibfnamefont{M.}~\bibnamefont{Stamenova}},
  \bibinfo{author}{\bibfnamefont{S.}~\bibnamefont{Sanvito}},
  \bibinfo{author}{\bibfnamefont{U.}~\bibnamefont{Bovensiepen}},
  \bibnamefont{et~al.}, \bibinfo{journal}{Phys. Rev. Lett.}
  \textbf{\bibinfo{volume}{119}}, \bibinfo{pages}{17202}
  (\bibinfo{year}{2017}).

\bibitem[{\citenamefont{Baida et~al.}(2011)\citenamefont{Baida, Mongin,
  Christofilos, Bachelier, Crut, Maioli, {Del Fatti}, and
  Vall{\'{e}}e}}]{Baida2011}
\bibinfo{author}{\bibfnamefont{H.}~\bibnamefont{Baida}},
  \bibinfo{author}{\bibfnamefont{D.}~\bibnamefont{Mongin}},
  \bibinfo{author}{\bibfnamefont{D.}~\bibnamefont{Christofilos}},
  \bibinfo{author}{\bibfnamefont{G.}~\bibnamefont{Bachelier}},
  \bibinfo{author}{\bibfnamefont{A.}~\bibnamefont{Crut}},
  \bibinfo{author}{\bibfnamefont{P.}~\bibnamefont{Maioli}},
  \bibinfo{author}{\bibfnamefont{N.}~\bibnamefont{{Del Fatti}}},
  \bibnamefont{and}
  \bibinfo{author}{\bibfnamefont{F.}~\bibnamefont{Vall{\'{e}}e}},
  \bibinfo{journal}{Phys. Rev. Lett.} \textbf{\bibinfo{volume}{107}},
  \bibinfo{pages}{57402} (\bibinfo{year}{2011}), ISSN \bibinfo{issn}{00319007}.

\bibitem[{\citenamefont{Mukherjee et~al.}(2013)\citenamefont{Mukherjee,
  Libisch, Large, Neumann, Brown, Cheng, Lassiter, Carter, Nordlander, and
  Halas}}]{Mukherjee2013}
\bibinfo{author}{\bibfnamefont{S.}~\bibnamefont{Mukherjee}},
  \bibinfo{author}{\bibfnamefont{F.}~\bibnamefont{Libisch}},
  \bibinfo{author}{\bibfnamefont{N.}~\bibnamefont{Large}},
  \bibinfo{author}{\bibfnamefont{O.}~\bibnamefont{Neumann}},
  \bibinfo{author}{\bibfnamefont{L.~V.} \bibnamefont{Brown}},
  \bibinfo{author}{\bibfnamefont{J.}~\bibnamefont{Cheng}},
  \bibinfo{author}{\bibfnamefont{J.~B.} \bibnamefont{Lassiter}},
  \bibinfo{author}{\bibfnamefont{E.~A.} \bibnamefont{Carter}},
  \bibinfo{author}{\bibfnamefont{P.}~\bibnamefont{Nordlander}},
  \bibnamefont{and} \bibinfo{author}{\bibfnamefont{N.~J.} \bibnamefont{Halas}},
  \bibinfo{journal}{Nano Lett.} \textbf{\bibinfo{volume}{13}},
  \bibinfo{pages}{240} (\bibinfo{year}{2013}), ISSN \bibinfo{issn}{15306984}.

\bibitem[{\citenamefont{Reiner et~al.}(2017)\citenamefont{Reiner, Nayak,
  Avraham, Norris, Yan, Fulga, Kang, Karzig, Shtrikman, and
  Beidenkopf}}]{Reiner2017}
\bibinfo{author}{\bibfnamefont{J.}~\bibnamefont{Reiner}},
  \bibinfo{author}{\bibfnamefont{A.~K.} \bibnamefont{Nayak}},
  \bibinfo{author}{\bibfnamefont{N.}~\bibnamefont{Avraham}},
  \bibinfo{author}{\bibfnamefont{A.}~\bibnamefont{Norris}},
  \bibinfo{author}{\bibfnamefont{B.}~\bibnamefont{Yan}},
  \bibinfo{author}{\bibfnamefont{I.~C.} \bibnamefont{Fulga}},
  \bibinfo{author}{\bibfnamefont{J.~H.} \bibnamefont{Kang}},
  \bibinfo{author}{\bibfnamefont{T.}~\bibnamefont{Karzig}},
  \bibinfo{author}{\bibfnamefont{H.}~\bibnamefont{Shtrikman}},
  \bibnamefont{and}
  \bibinfo{author}{\bibfnamefont{H.}~\bibnamefont{Beidenkopf}},
  \bibinfo{journal}{Phys. Rev. X} \textbf{\bibinfo{volume}{7}},
  \bibinfo{pages}{21016} (\bibinfo{year}{2017}), ISSN \bibinfo{issn}{21603308}.

\bibitem[{\citenamefont{Hartland et~al.}(2017)\citenamefont{Hartland, Besteiro,
  Johns, and Govorov}}]{Hartland2017}
\bibinfo{author}{\bibfnamefont{G.~V.} \bibnamefont{Hartland}},
  \bibinfo{author}{\bibfnamefont{L.~V.} \bibnamefont{Besteiro}},
  \bibinfo{author}{\bibfnamefont{P.}~\bibnamefont{Johns}}, \bibnamefont{and}
  \bibinfo{author}{\bibfnamefont{A.~O.} \bibnamefont{Govorov}},
  \bibinfo{journal}{ACS Energy Lett.} \textbf{\bibinfo{volume}{2}},
  \bibinfo{pages}{1641} (\bibinfo{year}{2017}).

\bibitem[{\citenamefont{Koopmans et~al.}(2000)\citenamefont{Koopmans, {Van
  Kampen}, Kohlhepp, and {De Jonge}}}]{Koopmans2000}
\bibinfo{author}{\bibfnamefont{B.}~\bibnamefont{Koopmans}},
  \bibinfo{author}{\bibfnamefont{M.}~\bibnamefont{{Van Kampen}}},
  \bibinfo{author}{\bibfnamefont{J.~T.} \bibnamefont{Kohlhepp}},
  \bibnamefont{and} \bibinfo{author}{\bibfnamefont{W.~J.~M.} \bibnamefont{{De
  Jonge}}}, \bibinfo{journal}{Phys. Rev. Lett.} \textbf{\bibinfo{volume}{85}},
  \bibinfo{pages}{844} (\bibinfo{year}{2000}), ISSN \bibinfo{issn}{00319007}.

\bibitem[{\citenamefont{Guidoni et~al.}(2002)\citenamefont{Guidoni,
  Beaurepaire, and Bigot}}]{Guidoni2002}
\bibinfo{author}{\bibfnamefont{L.}~\bibnamefont{Guidoni}},
  \bibinfo{author}{\bibfnamefont{E.}~\bibnamefont{Beaurepaire}},
  \bibnamefont{and} \bibinfo{author}{\bibfnamefont{J.~Y.} \bibnamefont{Bigot}},
  \bibinfo{journal}{Phys. Rev. Lett.} \textbf{\bibinfo{volume}{89}},
  \bibinfo{pages}{17401} (\bibinfo{year}{2002}), ISSN \bibinfo{issn}{10797114}.

\bibitem[{\citenamefont{McGroddy et~al.}(1965)\citenamefont{McGroddy,
  McAlister, and Stern}}]{McGroddy1965}
\bibinfo{author}{\bibfnamefont{J.~C.} \bibnamefont{McGroddy}},
  \bibinfo{author}{\bibfnamefont{A.~J.} \bibnamefont{McAlister}},
  \bibnamefont{and} \bibinfo{author}{\bibfnamefont{E.~A.} \bibnamefont{Stern}},
  \bibinfo{journal}{Phys. Rev.} \textbf{\bibinfo{volume}{139}},
  \bibinfo{pages}{A1844} (\bibinfo{year}{1965}), ISSN \bibinfo{issn}{0031899X}.

\bibitem[{\citenamefont{Haefner et~al.}(1994)\citenamefont{Haefner, Luck, and
  Mohler}}]{Haefner1994}
\bibinfo{author}{\bibfnamefont{P.}~\bibnamefont{Haefner}},
  \bibinfo{author}{\bibfnamefont{E.}~\bibnamefont{Luck}}, \bibnamefont{and}
  \bibinfo{author}{\bibfnamefont{E.}~\bibnamefont{Mohler}},
  \bibinfo{journal}{Phys. Status Solidi (B)} \textbf{\bibinfo{volume}{185}},
  \bibinfo{pages}{289} (\bibinfo{year}{1994}), ISSN \bibinfo{issn}{15213951}.

\bibitem[{\citenamefont{Etchegoin et~al.}(2006)\citenamefont{Etchegoin, {Le
  Ru}, and Meyer}}]{Etchegoin2006}
\bibinfo{author}{\bibfnamefont{P.~G.} \bibnamefont{Etchegoin}},
  \bibinfo{author}{\bibfnamefont{E.~C.} \bibnamefont{{Le Ru}}},
  \bibnamefont{and} \bibinfo{author}{\bibfnamefont{M.}~\bibnamefont{Meyer}},
  \bibinfo{journal}{J. Chem. Phys.} \textbf{\bibinfo{volume}{125}},
  \bibinfo{pages}{164705} (\bibinfo{year}{2006}), ISSN
  \bibinfo{issn}{00219606}.

\bibitem[{\citenamefont{Choi and Cahill}(2014)}]{Choi2014}
\bibinfo{author}{\bibfnamefont{G.-M.} \bibnamefont{Choi}} \bibnamefont{and}
  \bibinfo{author}{\bibfnamefont{D.~G.} \bibnamefont{Cahill}},
  \bibinfo{journal}{Phys. Rev. B} \textbf{\bibinfo{volume}{90}},
  \bibinfo{pages}{214432} (\bibinfo{year}{2014}).

\bibitem[{\citenamefont{Kimling et~al.}(2017)\citenamefont{Kimling, Choi,
  Brangham, Matalla-Wagner, Huebner, Kuschel, Yang, and Cahill}}]{Kimling2017}
\bibinfo{author}{\bibfnamefont{J.}~\bibnamefont{Kimling}},
  \bibinfo{author}{\bibfnamefont{G.~M.} \bibnamefont{Choi}},
  \bibinfo{author}{\bibfnamefont{J.~T.} \bibnamefont{Brangham}},
  \bibinfo{author}{\bibfnamefont{T.}~\bibnamefont{Matalla-Wagner}},
  \bibinfo{author}{\bibfnamefont{T.}~\bibnamefont{Huebner}},
  \bibinfo{author}{\bibfnamefont{T.}~\bibnamefont{Kuschel}},
  \bibinfo{author}{\bibfnamefont{F.}~\bibnamefont{Yang}}, \bibnamefont{and}
  \bibinfo{author}{\bibfnamefont{D.~G.} \bibnamefont{Cahill}},
  \bibinfo{journal}{Phys. Rev. Lett.} \textbf{\bibinfo{volume}{118}},
  \bibinfo{pages}{57201} (\bibinfo{year}{2017}).

\bibitem[{\citenamefont{Rudolf et~al.}(2012)\citenamefont{Rudolf,
  La-O-Vorakiat, Battiato, Adam, Shaw, Turgut, Maldonado, Mathias, Grychtol,
  Nembach et~al.}}]{Rudolf2012}
\bibinfo{author}{\bibfnamefont{D.}~\bibnamefont{Rudolf}},
  \bibinfo{author}{\bibfnamefont{C.}~\bibnamefont{La-O-Vorakiat}},
  \bibinfo{author}{\bibfnamefont{M.}~\bibnamefont{Battiato}},
  \bibinfo{author}{\bibfnamefont{R.}~\bibnamefont{Adam}},
  \bibinfo{author}{\bibfnamefont{J.~M.} \bibnamefont{Shaw}},
  \bibinfo{author}{\bibfnamefont{E.}~\bibnamefont{Turgut}},
  \bibinfo{author}{\bibfnamefont{P.}~\bibnamefont{Maldonado}},
  \bibinfo{author}{\bibfnamefont{S.}~\bibnamefont{Mathias}},
  \bibinfo{author}{\bibfnamefont{P.}~\bibnamefont{Grychtol}},
  \bibinfo{author}{\bibfnamefont{H.~T.} \bibnamefont{Nembach}},
  \bibnamefont{et~al.}, \bibinfo{journal}{Nat. Commun.}
  \textbf{\bibinfo{volume}{3}}, \bibinfo{pages}{1037} (\bibinfo{year}{2012}),
  ISSN \bibinfo{issn}{2041-1723}.

\bibitem[{\citenamefont{Bergeard et~al.}(2016)\citenamefont{Bergeard, Hehn,
  Mangin, Lengaigne, Montaigne, Lalieu, Koopmans, and
  Malinowski}}]{Bergeard2016}
\bibinfo{author}{\bibfnamefont{N.}~\bibnamefont{Bergeard}},
  \bibinfo{author}{\bibfnamefont{M.}~\bibnamefont{Hehn}},
  \bibinfo{author}{\bibfnamefont{S.}~\bibnamefont{Mangin}},
  \bibinfo{author}{\bibfnamefont{G.}~\bibnamefont{Lengaigne}},
  \bibinfo{author}{\bibfnamefont{F.}~\bibnamefont{Montaigne}},
  \bibinfo{author}{\bibfnamefont{M.~L.} \bibnamefont{Lalieu}},
  \bibinfo{author}{\bibfnamefont{B.}~\bibnamefont{Koopmans}}, \bibnamefont{and}
  \bibinfo{author}{\bibfnamefont{G.}~\bibnamefont{Malinowski}},
  \bibinfo{journal}{Phys. Rev. Lett.} \textbf{\bibinfo{volume}{117}},
  \bibinfo{pages}{147203} (\bibinfo{year}{2016}), ISSN
  \bibinfo{issn}{10797114}.

\bibitem[{\citenamefont{Banerjee et~al.}(2016)\citenamefont{Banerjee, Pal,
  Ahlberg, Nguyen, {\AA}kerman, and Barman}}]{Banerjee2016}
\bibinfo{author}{\bibfnamefont{C.}~\bibnamefont{Banerjee}},
  \bibinfo{author}{\bibfnamefont{S.}~\bibnamefont{Pal}},
  \bibinfo{author}{\bibfnamefont{M.}~\bibnamefont{Ahlberg}},
  \bibinfo{author}{\bibfnamefont{T.~N.} \bibnamefont{Nguyen}},
  \bibinfo{author}{\bibfnamefont{J.}~\bibnamefont{{\AA}kerman}},
  \bibnamefont{and} \bibinfo{author}{\bibfnamefont{A.}~\bibnamefont{Barman}},
  \bibinfo{journal}{RSC Advances} \textbf{\bibinfo{volume}{6}},
  \bibinfo{pages}{80168} (\bibinfo{year}{2016}), ISSN \bibinfo{issn}{20462069}.

\bibitem[{\citenamefont{Hofherr et~al.}(2017)\citenamefont{Hofherr, Maldonado,
  Schmitt, Berritta, Bierbrauer, Sadashivaiah, Schellekens, Koopmans, Steil,
  Cinchetti et~al.}}]{Hofherr2017}
\bibinfo{author}{\bibfnamefont{M.}~\bibnamefont{Hofherr}},
  \bibinfo{author}{\bibfnamefont{P.}~\bibnamefont{Maldonado}},
  \bibinfo{author}{\bibfnamefont{O.}~\bibnamefont{Schmitt}},
  \bibinfo{author}{\bibfnamefont{M.}~\bibnamefont{Berritta}},
  \bibinfo{author}{\bibfnamefont{U.}~\bibnamefont{Bierbrauer}},
  \bibinfo{author}{\bibfnamefont{S.}~\bibnamefont{Sadashivaiah}},
  \bibinfo{author}{\bibfnamefont{A.~J.} \bibnamefont{Schellekens}},
  \bibinfo{author}{\bibfnamefont{B.}~\bibnamefont{Koopmans}},
  \bibinfo{author}{\bibfnamefont{D.}~\bibnamefont{Steil}},
  \bibinfo{author}{\bibfnamefont{M.}~\bibnamefont{Cinchetti}},
  \bibnamefont{et~al.}, \bibinfo{journal}{Phys. Rev. B}
  \textbf{\bibinfo{volume}{96}}, \bibinfo{pages}{100403}
  (\bibinfo{year}{2017}),
  \urlprefix\url{https://link.aps.org/doi/10.1103/PhysRevB.96.100403}.

\bibitem[{\citenamefont{Zvezdin and Kotov}(1997)}]{Zvezdin1997}
\bibinfo{author}{\bibfnamefont{A.~K.} \bibnamefont{Zvezdin}} \bibnamefont{and}
  \bibinfo{author}{\bibfnamefont{V.~A.} \bibnamefont{Kotov}},
  \emph{\bibinfo{title}{{Modern magnetooptics and magnetooptical materials}}}
  (\bibinfo{publisher}{CRC Press}, \bibinfo{year}{1997}), ISBN
  \bibinfo{isbn}{075030362X}.

\bibitem[{\citenamefont{Oppeneer}(2001)}]{Oppeneer2001}
\bibinfo{author}{\bibfnamefont{P.}~\bibnamefont{Oppeneer}}, in
  \emph{\bibinfo{booktitle}{Handbook of Magnetic Materials}}
  (\bibinfo{publisher}{Elsevier}, \bibinfo{year}{2001}),
  vol.~\bibinfo{volume}{13}, pp. \bibinfo{pages}{229--422}, ISBN
  \bibinfo{isbn}{1567-2719}.

\bibitem[{\citenamefont{Bennemann}(1998)}]{Bennemann1998}
\bibinfo{author}{\bibfnamefont{K.~H.} \bibnamefont{Bennemann}},
  \emph{\bibinfo{title}{{Nonlinear optics in metals}}}
  (\bibinfo{publisher}{Clarendon Press}, \bibinfo{address}{Oxford},
  \bibinfo{year}{1998}), ISBN \bibinfo{isbn}{9780198518938}.

\bibitem[{\citenamefont{Voigt}(1908)}]{Voigt1915}
\bibinfo{author}{\bibfnamefont{W.}~\bibnamefont{Voigt}}, in
  \emph{\bibinfo{booktitle}{Magneto und Elektrooptik (Teubner, Leipzig)}}
  (\bibinfo{publisher}{Teubner, Leipzig}, \bibinfo{year}{1908}), p.
  \bibinfo{pages}{393}.

\bibitem[{\citenamefont{Zak et~al.}(1990)\citenamefont{Zak, Moog, Liu, and
  Bader}}]{Zak1990}
\bibinfo{author}{\bibfnamefont{J.}~\bibnamefont{Zak}},
  \bibinfo{author}{\bibfnamefont{E.~R.} \bibnamefont{Moog}},
  \bibinfo{author}{\bibfnamefont{C.}~\bibnamefont{Liu}}, \bibnamefont{and}
  \bibinfo{author}{\bibfnamefont{S.~D.} \bibnamefont{Bader}},
  \bibinfo{journal}{J. Magn. Magn. Mater.} \textbf{\bibinfo{volume}{89}},
  \bibinfo{pages}{107} (\bibinfo{year}{1990}), ISSN \bibinfo{issn}{0304-8853}.

\bibitem[{\citenamefont{Johnson and Christy}(1972)}]{Johnson1972}
\bibinfo{author}{\bibfnamefont{P.~B.} \bibnamefont{Johnson}} \bibnamefont{and}
  \bibinfo{author}{\bibfnamefont{R.~W.} \bibnamefont{Christy}},
  \bibinfo{journal}{Phys. Rev. B} \textbf{\bibinfo{volume}{6}},
  \bibinfo{pages}{4370} (\bibinfo{year}{1972}), ISSN \bibinfo{issn}{01631829}.

\bibitem[{\citenamefont{Pan et~al.}(1989)\citenamefont{Pan, Wei, and
  Shen}}]{Pan1989}
\bibinfo{author}{\bibfnamefont{R.~P.} \bibnamefont{Pan}},
  \bibinfo{author}{\bibfnamefont{H.~D.} \bibnamefont{Wei}}, \bibnamefont{and}
  \bibinfo{author}{\bibfnamefont{Y.~R.} \bibnamefont{Shen}},
  \bibinfo{journal}{Phys. Rev. B} \textbf{\bibinfo{volume}{39}},
  \bibinfo{pages}{1229} (\bibinfo{year}{1989}), ISSN \bibinfo{issn}{01631829}.

\bibitem[{\citenamefont{M{\"{u}}hge et~al.}(1994)\citenamefont{M{\"{u}}hge,
  Stierle, Metoki, Zabel, and Pietsch}}]{Muhge1994}
\bibinfo{author}{\bibfnamefont{T.}~\bibnamefont{M{\"{u}}hge}},
  \bibinfo{author}{\bibfnamefont{A.}~\bibnamefont{Stierle}},
  \bibinfo{author}{\bibfnamefont{N.}~\bibnamefont{Metoki}},
  \bibinfo{author}{\bibfnamefont{H.}~\bibnamefont{Zabel}}, \bibnamefont{and}
  \bibinfo{author}{\bibfnamefont{U.}~\bibnamefont{Pietsch}},
  \bibinfo{journal}{Appl. Phys. A} \textbf{\bibinfo{volume}{59}},
  \bibinfo{pages}{659} (\bibinfo{year}{1994}), ISSN \bibinfo{issn}{1432-0630}.

\bibitem[{\citenamefont{Razdolski
  et~al.}(2017{\natexlab{b}})\citenamefont{Razdolski, Alekhin, Ilin, Meyburg,
  Roddatis, Diesing, Bovensiepen, and Melnikov}}]{Razdolski2017a}
\bibinfo{author}{\bibfnamefont{I.}~\bibnamefont{Razdolski}},
  \bibinfo{author}{\bibfnamefont{A.}~\bibnamefont{Alekhin}},
  \bibinfo{author}{\bibfnamefont{N.}~\bibnamefont{Ilin}},
  \bibinfo{author}{\bibfnamefont{J.~P.} \bibnamefont{Meyburg}},
  \bibinfo{author}{\bibfnamefont{V.}~\bibnamefont{Roddatis}},
  \bibinfo{author}{\bibfnamefont{D.}~\bibnamefont{Diesing}},
  \bibinfo{author}{\bibfnamefont{U.}~\bibnamefont{Bovensiepen}},
  \bibnamefont{and} \bibinfo{author}{\bibfnamefont{A.}~\bibnamefont{Melnikov}},
  \bibinfo{journal}{Nat. Commun.} \textbf{\bibinfo{volume}{8}},
  \bibinfo{pages}{15007} (\bibinfo{year}{2017}{\natexlab{b}}).

\bibitem[{\citenamefont{Krinchik and Artem'ev}(1968)}]{Krinchik1968}
\bibinfo{author}{\bibfnamefont{G.~S.} \bibnamefont{Krinchik}} \bibnamefont{and}
  \bibinfo{author}{\bibfnamefont{a.~V.} \bibnamefont{Artem'ev}},
  \bibinfo{journal}{Soviet Physics JETP} \textbf{\bibinfo{volume}{26}},
  \bibinfo{pages}{1080} (\bibinfo{year}{1968}).

\bibitem[{\citenamefont{Brorson et~al.}(1987)\citenamefont{Brorson, Fujimoto,
  and Ippen}}]{Brorson1987}
\bibinfo{author}{\bibfnamefont{S.~D.} \bibnamefont{Brorson}},
  \bibinfo{author}{\bibfnamefont{J.~G.} \bibnamefont{Fujimoto}},
  \bibnamefont{and} \bibinfo{author}{\bibfnamefont{E.~P.} \bibnamefont{Ippen}},
  \bibinfo{journal}{Phys. Rev. Lett.} \textbf{\bibinfo{volume}{59}},
  \bibinfo{pages}{1962} (\bibinfo{year}{1987}).

\bibitem[{\citenamefont{Zhukov et~al.}(2006)\citenamefont{Zhukov, Chulkov, and
  Echenique}}]{Zhukov2006}
\bibinfo{author}{\bibfnamefont{V.~P.} \bibnamefont{Zhukov}},
  \bibinfo{author}{\bibfnamefont{E.~V.} \bibnamefont{Chulkov}},
  \bibnamefont{and} \bibinfo{author}{\bibfnamefont{P.~M.}
  \bibnamefont{Echenique}}, \bibinfo{journal}{Phys. Rev. B}
  \textbf{\bibinfo{volume}{73}}, \bibinfo{pages}{125105}
  (\bibinfo{year}{2006}).

\bibitem[{\citenamefont{Oppeneer et~al.}(1992)\citenamefont{Oppeneer, Maurer,
  Sticht, and Kubler}}]{Oppeneer1992}
\bibinfo{author}{\bibfnamefont{P.~M.} \bibnamefont{Oppeneer}},
  \bibinfo{author}{\bibfnamefont{T.}~\bibnamefont{Maurer}},
  \bibinfo{author}{\bibfnamefont{J.}~\bibnamefont{Sticht}}, \bibnamefont{and}
  \bibinfo{author}{\bibfnamefont{J.}~\bibnamefont{Kubler}},
  \bibinfo{journal}{Phys. Rev. B} \textbf{\bibinfo{volume}{45}},
  \bibinfo{pages}{10924} (\bibinfo{year}{1992}), ISSN \bibinfo{issn}{01631829}.

\bibitem[{\citenamefont{McPeak et~al.}(2015)\citenamefont{McPeak, Jayanti,
  Kress, Meyer, Iotti, Rossinelli, and Norris}}]{mcpeak2015}
\bibinfo{author}{\bibfnamefont{K.~M.} \bibnamefont{McPeak}},
  \bibinfo{author}{\bibfnamefont{S.~V.} \bibnamefont{Jayanti}},
  \bibinfo{author}{\bibfnamefont{S.~J.~P.} \bibnamefont{Kress}},
  \bibinfo{author}{\bibfnamefont{S.}~\bibnamefont{Meyer}},
  \bibinfo{author}{\bibfnamefont{S.}~\bibnamefont{Iotti}},
  \bibinfo{author}{\bibfnamefont{A.}~\bibnamefont{Rossinelli}},
  \bibnamefont{and} \bibinfo{author}{\bibfnamefont{D.~J.}
  \bibnamefont{Norris}}, \bibinfo{journal}{ACS Photonics}
  \textbf{\bibinfo{volume}{2}}, \bibinfo{pages}{326} (\bibinfo{year}{2015}),
  \urlprefix\url{https://doi.org/10.1021/ph5004237}.

\bibitem[{\citenamefont{Babar and Weaver}(2015)}]{Babar2015}
\bibinfo{author}{\bibfnamefont{S.}~\bibnamefont{Babar}} \bibnamefont{and}
  \bibinfo{author}{\bibfnamefont{J.~H.} \bibnamefont{Weaver}},
  \bibinfo{journal}{Appl. Opt.} \textbf{\bibinfo{volume}{54}},
  \bibinfo{pages}{477} (\bibinfo{year}{2015}),
  \urlprefix\url{http://ao.osa.org/abstract.cfm?URI=ao-54-3-477}.

\bibitem[{\citenamefont{Olmon et~al.}(2012)\citenamefont{Olmon, Slovick,
  Johnson, Shelton, Oh, Boreman, and Raschke}}]{Olmon2012}
\bibinfo{author}{\bibfnamefont{R.~L.} \bibnamefont{Olmon}},
  \bibinfo{author}{\bibfnamefont{B.}~\bibnamefont{Slovick}},
  \bibinfo{author}{\bibfnamefont{T.~W.} \bibnamefont{Johnson}},
  \bibinfo{author}{\bibfnamefont{D.}~\bibnamefont{Shelton}},
  \bibinfo{author}{\bibfnamefont{S.-H.} \bibnamefont{Oh}},
  \bibinfo{author}{\bibfnamefont{G.~D.} \bibnamefont{Boreman}},
  \bibnamefont{and} \bibinfo{author}{\bibfnamefont{M.~B.}
  \bibnamefont{Raschke}}, \bibinfo{journal}{Phys. Rev. B}
  \textbf{\bibinfo{volume}{86}}, \bibinfo{pages}{235147}
  (\bibinfo{year}{2012}),
  \urlprefix\url{https://link.aps.org/doi/10.1103/PhysRevB.86.235147}.

\bibitem[{\citenamefont{You and Shin}(1996)}]{You1996}
\bibinfo{author}{\bibfnamefont{C.-Y.} \bibnamefont{You}} \bibnamefont{and}
  \bibinfo{author}{\bibfnamefont{S.-C.} \bibnamefont{Shin}},
  \bibinfo{journal}{Appl. Phys. Lett.} \textbf{\bibinfo{volume}{69}},
  \bibinfo{pages}{1315} (\bibinfo{year}{1996}), ISSN \bibinfo{issn}{0003-6951}.

\bibitem[{\citenamefont{Zeman and Schatz}(1987)}]{Zeman1987}
\bibinfo{author}{\bibfnamefont{E.~J.} \bibnamefont{Zeman}} \bibnamefont{and}
  \bibinfo{author}{\bibfnamefont{G.~C.} \bibnamefont{Schatz}},
  \bibinfo{journal}{J. Phys. Chem.} \textbf{\bibinfo{volume}{91}},
  \bibinfo{pages}{634} (\bibinfo{year}{1987}), ISSN \bibinfo{issn}{0022-3654}.

\bibitem[{\citenamefont{Haynes}(2016)}]{Haynes2016}
\bibinfo{author}{\bibfnamefont{W.~M.} \bibnamefont{Haynes}},
  \emph{\bibinfo{title}{CRC Handbook of Chemistry and Physics}}
  (\bibinfo{publisher}{CRC Press}, \bibinfo{year}{2016}), ISBN
  \bibinfo{isbn}{9781498754293},
  \urlprefix\url{https://books.google.se/books?id=VVezDAAAQBAJ}.

\bibitem[{\citenamefont{Kruglyak et~al.}(2005)\citenamefont{Kruglyak, Hicken,
  Ali, Hickey, Pym, and Tanner}}]{Kruglyak2005}
\bibinfo{author}{\bibfnamefont{V.~V.} \bibnamefont{Kruglyak}},
  \bibinfo{author}{\bibfnamefont{R.~J.} \bibnamefont{Hicken}},
  \bibinfo{author}{\bibfnamefont{M.}~\bibnamefont{Ali}},
  \bibinfo{author}{\bibfnamefont{B.~J.} \bibnamefont{Hickey}},
  \bibinfo{author}{\bibfnamefont{A.~T.~G.} \bibnamefont{Pym}},
  \bibnamefont{and} \bibinfo{author}{\bibfnamefont{B.~K.}
  \bibnamefont{Tanner}}, \bibinfo{journal}{Phys. Rev. B}
  \textbf{\bibinfo{volume}{71}}, \bibinfo{pages}{233104}
  (\bibinfo{year}{2005}),
  \urlprefix\url{https://link.aps.org/doi/10.1103/PhysRevB.71.233104}.

\bibitem[{\citenamefont{Cooper et~al.}(1965)\citenamefont{Cooper, Ehrenreich,
  and Philipp}}]{Cooper1965}
\bibinfo{author}{\bibfnamefont{B.~R.} \bibnamefont{Cooper}},
  \bibinfo{author}{\bibfnamefont{H.}~\bibnamefont{Ehrenreich}},
  \bibnamefont{and} \bibinfo{author}{\bibfnamefont{H.~R.}
  \bibnamefont{Philipp}}, \bibinfo{journal}{Phys. Rev.}
  \textbf{\bibinfo{volume}{138}}, \bibinfo{pages}{A494} (\bibinfo{year}{1965}),
  ISSN \bibinfo{issn}{0031899X}.

\bibitem[{\citenamefont{Sep{\'{u}}lveda
  et~al.}(2010)\citenamefont{Sep{\'{u}}lveda, Gonz{\'{a}}lez-D{\'{i}}az,
  Garc{\'{i}}a-Mart{\'{i}}n, Lechuga, and Armelles}}]{Sepulveda2010}
\bibinfo{author}{\bibfnamefont{B.}~\bibnamefont{Sep{\'{u}}lveda}},
  \bibinfo{author}{\bibfnamefont{J.~B.}
  \bibnamefont{Gonz{\'{a}}lez-D{\'{i}}az}},
  \bibinfo{author}{\bibfnamefont{A.}~\bibnamefont{Garc{\'{i}}a-Mart{\'{i}}n}},
  \bibinfo{author}{\bibfnamefont{L.~M.} \bibnamefont{Lechuga}},
  \bibnamefont{and} \bibinfo{author}{\bibfnamefont{G.}~\bibnamefont{Armelles}},
  \bibinfo{journal}{Phys. Rev. Lett.} \textbf{\bibinfo{volume}{104}},
  \bibinfo{pages}{147401} (\bibinfo{year}{2010}), ISSN
  \bibinfo{issn}{00319007}.

\bibitem[{\citenamefont{Th{\`{e}}ye}(1970)}]{Theye1970}
\bibinfo{author}{\bibfnamefont{M.~L.} \bibnamefont{Th{\`{e}}ye}},
  \bibinfo{journal}{Phys. Rev. B} \textbf{\bibinfo{volume}{2}},
  \bibinfo{pages}{3060} (\bibinfo{year}{1970}), ISSN \bibinfo{issn}{01631829}.

\bibitem[{\citenamefont{Erskine and Stern}(1973)}]{Erskine1973}
\bibinfo{author}{\bibfnamefont{J.~L.} \bibnamefont{Erskine}} \bibnamefont{and}
  \bibinfo{author}{\bibfnamefont{E.~A.} \bibnamefont{Stern}},
  \bibinfo{journal}{Phys. Rev. B} \textbf{\bibinfo{volume}{8}},
  \bibinfo{pages}{1239} (\bibinfo{year}{1973}), ISSN \bibinfo{issn}{01631829}.

\bibitem[{\citenamefont{Fivaz}(1969)}]{Fivaz1969}
\bibinfo{author}{\bibfnamefont{R.~C.} \bibnamefont{Fivaz}},
  \bibinfo{journal}{Phys. Rev.} \textbf{\bibinfo{volume}{183}},
  \bibinfo{pages}{586} (\bibinfo{year}{1969}), ISSN \bibinfo{issn}{0031899X},
  \urlprefix\url{https://link.aps.org/doi/10.1103/PhysRev.183.586}.

\bibitem[{\citenamefont{Elezzabi et~al.}(1996)\citenamefont{Elezzabi, Freeman,
  and Johnson}}]{Elezzabi1996}
\bibinfo{author}{\bibfnamefont{A.~Y.} \bibnamefont{Elezzabi}},
  \bibinfo{author}{\bibfnamefont{M.~R.} \bibnamefont{Freeman}},
  \bibnamefont{and} \bibinfo{author}{\bibfnamefont{M.}~\bibnamefont{Johnson}},
  \bibinfo{journal}{Phys. Rev. Lett.} \textbf{\bibinfo{volume}{77}},
  \bibinfo{pages}{3220} (\bibinfo{year}{1996}), ISSN \bibinfo{issn}{10797114}.

\bibitem[{\citenamefont{Pershan}(1963)}]{Pershan1963}
\bibinfo{author}{\bibfnamefont{P.~S.} \bibnamefont{Pershan}},
  \bibinfo{journal}{Phys. Rev.} \textbf{\bibinfo{volume}{130}},
  \bibinfo{pages}{919} (\bibinfo{year}{1963}).

\end{thebibliography}

\end{document}